\documentclass[fleqn,usenatbib]{mnras}

\usepackage{newtxtext,newtxmath}
\usepackage[T1]{fontenc}
\usepackage{ae,aecompl}

\usepackage{graphicx}	
\usepackage{amsmath}	
\usepackage{amssymb}	

\usepackage{lipsum}

\usepackage{multirow}

\newcommand\Tstrut{\rule{0pt}{2.6ex}}         

\usepackage{xcolor}

\newif\ifref
\reftrue 
\reffalse
\definecolor{darkred}{rgb}{0.7, 0, 0}
\newcommand{\mb}[1]{\ifref\textcolor{darkred}{#1}\else #1\fi}

\newif\ifreff
\refftrue 
\refffalse
\definecolor{darkred}{rgb}{0.7, 0, 0}
\newcommand{\mbb}[1]{\ifreff\textcolor{darkred}{#1}\else #1\fi}

\title[Fundamental parameters of Cepheid and RR~Lyrae stars from light curve structure]{When a Period Is Not a Full Stop: Light Curve Structure Reveals Fundamental Parameters of Cepheid and RR~Lyrae Stars}

\author[E.\ P.\ Bellinger et al.]{
Earl P.\ Bellinger,$^{1}$\thanks{\href{mailto:bellinger@phys.au.dk}{bellinger@phys.au.dk}}\thanks{SAC Fellow}
Shashi M.\ Kanbur,$^{2}$
Anupam Bhardwaj,$^{3}$\thanks{KIAA Fellow} and 
\newauthor\hphantom{ }Marcella Marconi$^{4}$
\\
$^{1}$Stellar Astrophysics Centre, Department of Physics and Astronomy, Aarhus University, Denmark\\
$^{2}$State University of New York, Oswego, NY 13126, USA\\
$^{3}$Kavli Institute for Astronomy and Astrophysics, Peking University, Yi He Yuan Lu 5, Hai Dian District, Beijing 100871, China\\
$^{4}$INAF-Osservatorio Astronomico di Capodimonte, Salita Moiariello 16, 80131 Napoli, Italy
}

\date{Accepted XXX. Received YYY; in original form ZZZ}
\pubyear{2019}

\begin{document}
\label{firstpage}
\pagerange{\pageref{firstpage}--\pageref{lastpage}}
\maketitle

\begin{abstract}
The period of pulsation and the structure of the light curve for Cepheid and RR~Lyrae variables depend on the fundamental parameters of the star: mass, radius, luminosity, and effective temperature. 
Here we train artificial neural networks on theoretical pulsation models to predict the fundamental parameters of these stars based on their period and light curve structure. 
We find significant improvements to estimates of these parameters made using light curve structure and period over estimates made using only the period. 
Given that the models are able to reproduce most observables, we find that the fundamental parameters of these stars can be estimated up to 60\% more accurately when light curve structure is taken into consideration. 
We quantify which aspects of light curve structure are most important in determining fundamental parameters, and find for example that the second Fourier amplitude component of RR~Lyrae light curves is even more important than period in determining the effective temperature of the star. 
We apply this analysis to observations of hundreds Cepheids in the Large Magellanic Cloud and thousands of RR~Lyrae in the Magellanic Clouds and Galactic bulge to produce catalogs of estimated masses, radii, luminosities, and other parameters of these stars. 
As an example application, we estimate Wesenheit indices and use those to derive distance moduli to the Magellanic Clouds of $\mu_{\text{LMC},\text{CEP}} = 18.688 \pm 0.093$, $\mu_{\text{LMC},\text{RRL}} = 18.52 \pm 0.14$, and $\mu_{\text{SMC},\text{RRL}} = 18.88 \pm 0.17$~mag. 
\end{abstract}


\section{Introduction}
Cepheids and RR~Lyraes are evolved stars whose light curves exhibit large-amplitude oscillations with typical periods of days and hours, respectively.
Cepheids, thanks to their Period-Luminosity (PL) relation, have a crucial importance in the extra-galactic distance scale and in the comparison with Cosmic Microwave Background-independent estimates of the Hubble constant \citep{freedman2001,2019ApJ...876...85R}. 
RR~Lyraes are excellent tracers of old stellar populations in the Milky Way and the Magellanic Clouds \citep[e.g.,][]{2015ApJ...811..113P, 2018MNRAS.481.1195M}, and follow a strong PL relation at infrared wavelengths that can be used to formulate a Population II distance scale \citep{beaton2016}. 
In recent years, multi-wavelength Cepheid and RR~Lyrae light curves have become available from various ground and space-based microlensing and extra-solar planet hunting projects, such as OGLE-III/IV, \emph{Kepler}, Gaia, and TESS. 

Characterizing the non-linear structure of Cepheid and RR~Lyrae light curves began for the most part with the work of \citet{1981ApJ...248..291S}, who studied Galactic Cepheids using Fourier analysis and were able to show distinct progressions with period amongst the Fourier coefficients. Later \citet{1981PASP...93..550S} extended the same method to RR~Lyraes. Since then Fourier decomposition of radially pulsating Cepheid and RR~Lyraes has become standard. \citet{1983ApJ...266..787S} were among the first to compare theoretical and observed light curves (in this case for classical Cepheids) using Fourier decomposition. \citet{1997A&A...322..817F} and \citet{1998MNRAS.295..834K} amongst others compared theoretical model light curves of RR~Lyraes to observations. These studies found, for example, that Fourier methods could help understanding the Hertzsprung progression of the so called ``bump Cepheids'' \citep[e.g.,][]{2002ApJ...565L..83B} and outline regions where the models agreed with observations. 
\citet{2017MNRAS.466.2805B} and \citet{2018MNRAS.481.2000D} describe the latest in this area for Cepheids and RR~Lyraes respectively. \citet{2017MNRAS.466.2805B} found that theoretical Fourier amplitude parameters are significantly larger at optical wavelengths when compared to observations, and that a higher convective efficiency parameter in the models can resolve this discrepancy. 

The Fourier decomposition entails fitting light curves (theoretical and observed alike) with a function of the form
\begin{equation} \label{eq:Fourier}
    M(t) = A_0 + \sum_{k=1}^{N} A_k \cos\left( 
            k{\omega}t + {\phi}_k
    \right)
\end{equation}
where $M$ is the magnitude, $t$ is the time of observation, and ${\omega=2\pi/P}$, with $P$ being the period of pulsation. The quantities
$A_0,\ldots A_N$ and $\phi_1,\ldots \phi_N$ describe the structure of the light curve and can be used to quantitatively compare between observed and theoretical light curves. 
In addition, one can consider derived features of the light curve, such as the skewness and acuteness. 
Skewness is the ratio of the phase duration of the descending branch to the rising branch \mb{(not to be confused with the definition of the skewness of a statistical distribution)}, while acuteness is the ratio of the phase duration when the magnitude is fainter than the median magnitude to when it is brighter than the median magnitude \citep{2000A&A...360..245B, 2015MNRAS.447.3342B}. 
In brief, skewness defines left/right asymmetry in the light curve, and acuteness defines top/down asymmetry. 
For a pure sinusoidal curve both values are unity. 

Additionally, there have been a number of attempts to connect light curve structure to internal physics, perhaps most notably being the connection of the Fourier parameter \mb{${{\phi}_{31} \equiv \phi_3 - 3\phi_1}$} to the metallicity of RR~Lyrae stars. Initially this was a model dependent approach \citep{1993ApJ...410..526S} applied to observational data, but was later developed into a completely empirical technique \citep{1996A&A...312..111J}. This latter relation has had several revisions and extensions \citep[e.g.,][]{2005AcA....55...59S, 2013MNRAS.434.2418K, 2013ApJ...773..181N}. 
Pulsation models have also been used to estimate global physical parameters in past work. 
For example, \citet{1997ApJ...485L..25W} \mb{performed} the first application of the light curve model fitting technique to a Classical Cepheid. 
Subsequently \citet{2002ApJ...565L..83B}, \citet{2002ApJ...578..144K, 2006ApJ...642..834K} and \citet{2013MNRAS.428.2185M} applied this technique to Large Magellanic Cloud (LMC) Cepheids, including the fitting of the radial velocity curve when available \citep{2013MNRAS.428.2185M}, while \citep{2008ApJ...674L..93N} applied the model fitting technique to the light, radial velocity and radius curve of $\delta$ Cephei. 
\citet{2013ApJ...768L...6M} estimated physical parameters for a Cepheid in a binary systems using a grid of models to match the observed light, radial velocity and radius curves. 
More recently, \citet{2017MNRAS.466.3206M} fit multiband infrared light and velocity curve observations of Cepheids in the Small Magellanic Cloud (SMC) with a grid of pulsation models to obtain a distance to the SMC that is consistent with the latest estimates. 
The same technique was also successfully applied to field and cluster RR Lyrae \citep[see, e.g.,][and references therein]{2005AJ....129.2257M, 2007A&A...474..557M}. 

In this work, we build on these studies in the following way. 
We make the following assumptions: given the mass $M$, luminosity $L$, effective temperature $T_{\text{eff}}$ and composition, we can determine the period and light curve structure through a forward modeling process, i.e., a stellar pulsation radiation hydrodynamics code, and that this forward modelling process is at least approximately correct. 
Given this, we study the extent to which the period and, in particular, light curve structure, can help to constrain these fundamental parameters $M,L,T_{\text{eff}}$ and parameters derived thereof, such as the stellar radius or absolute magnitudes in different passbands. 
We first look for connections between these parameters and the period and light curve structure {\it purely} amongst our model grid, assess the statistical significance of adding light curve structure measures to our set of independent variables, and then apply our relations to observed Cepheid and RR~Lyrae stars.

The layout of the paper is as follows. 
In Section~\ref{sec:data} we describe the adopted theoretical models and observational data. 
In Section~\ref{sec:methods} we describe the methodology of our machine learning-based approach and demonstrate on theoretical models that the use of information contained within the light curve structure significantly improves estimates of fundamental stellar parameters. 
In Section~\ref{sec:applications} we apply this analysis to stars in the Galactic bulge as well as the Magellanic Clouds and present estimates of the fundamental parameters of these stars. 
Finally, we discuss our results and finish with a discussion about possible next steps.

\section{Models and Data} \label{sec:data}
We use nonlinear 1D models of Cepheids and RR Lyraes that include a non-local, time-dependent theory of convection 
\citep[][and references therein]{1999ApJS..122..167B, 2013MNRAS.428.2185M, 2015ApJ...808...50M, 2018ApJ...864L..13M}. 
On this basis relations connecting the period to intrinsic stellar parameters such as  the luminosity, the effective temperature, the mass and the metallicity have been derived 
\citep[see, e.g.,][]{1999ApJ...512..711B, 2000A&A...354..610C, 2015ApJ...808...50M, 2015ApJ...799..165B}.
In this work we seek to extend these linear relations by looking for non-linear relations that incorporate measures of light curve structure.


In the Cepheid case, the model grid composition is appropriate for the LMC, i.e., $Y=0.25$ and $Z=0.008$. For each mass, the luminosity is given by the canonical mass--luminosity (ML) relation given by \citet{2000A&A...360..245B}. We also have models with a brighter luminosity (by 0.25~dex) level to account for mild overshooting. In total we use 391 Cepheid models. 

In the RR~Lyrae case, the models have metallicities ranging from $Z=0.0001$ to $Z=0.02$ and helium abundances ranging from $Y=0.245$ to $Y=0.27$. Each composition has several sets of masses and luminosities fixed according to core helium burning horizontal branch evolutionary models. In total we use 269 RR~Lyrae models. 

For both Cepheids and RR~Lyraes, the range of effective temperatures for a given ML pair covers the width of the instability strip. The theoretical bolometric light curves were converted into theoretical $V,I$ light curves using static model atmospheres by \citet{1997A&A...324..432C}. All models were followed until a stable fundamental model limit cycle was achieved. This procedure also resulted in a theoretical non-linear period for each model. 
\citet{2017MNRAS.466.2805B} and \citet{2018MNRAS.481.2000D} presented an analysis of the characteristics of the light curve structure for the models used in this study. 

\begin{figure}
    \centering
    \includegraphics[width=\linewidth,keepaspectratio]{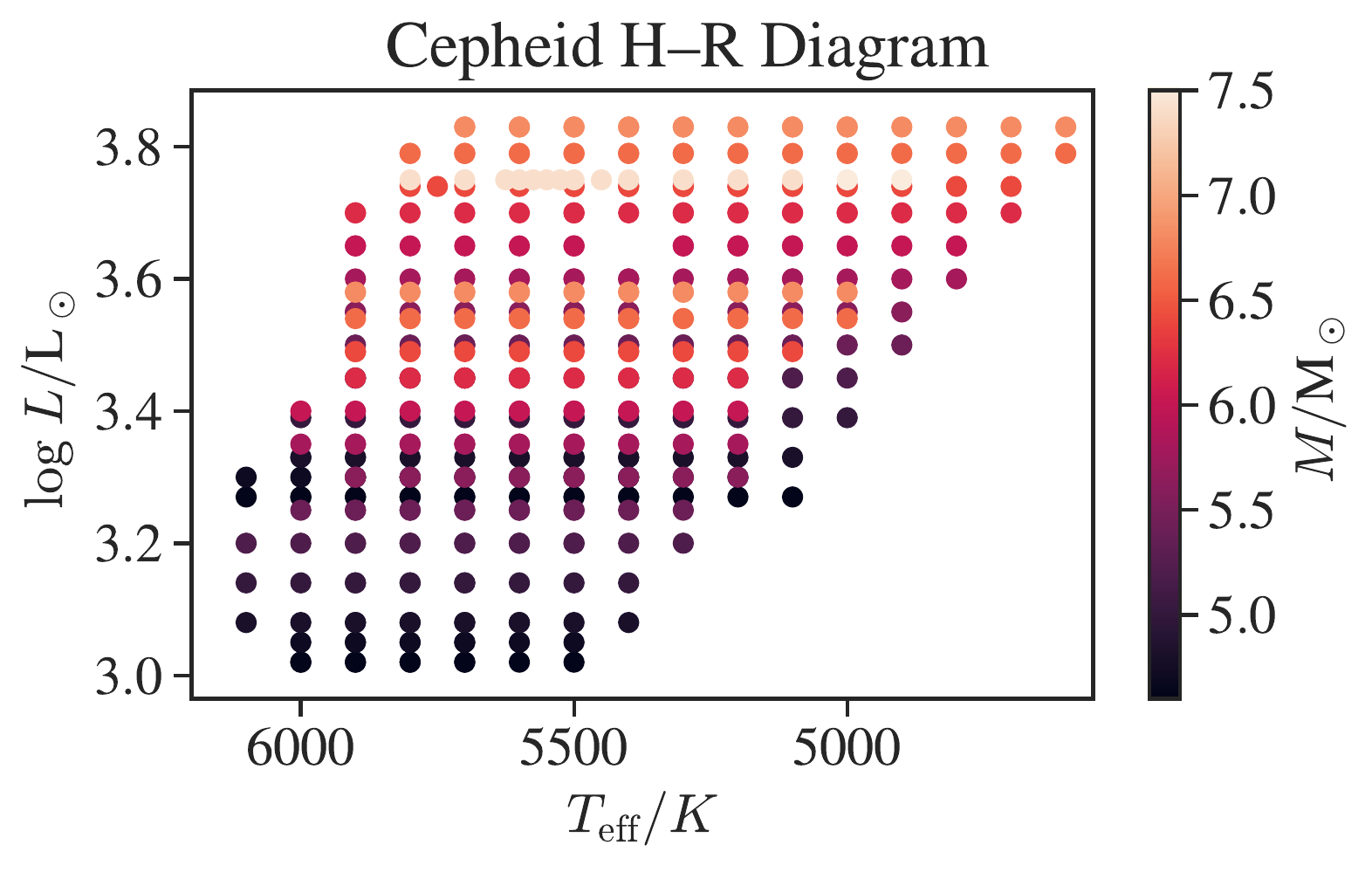}\\%
    \includegraphics[width=\linewidth,keepaspectratio]{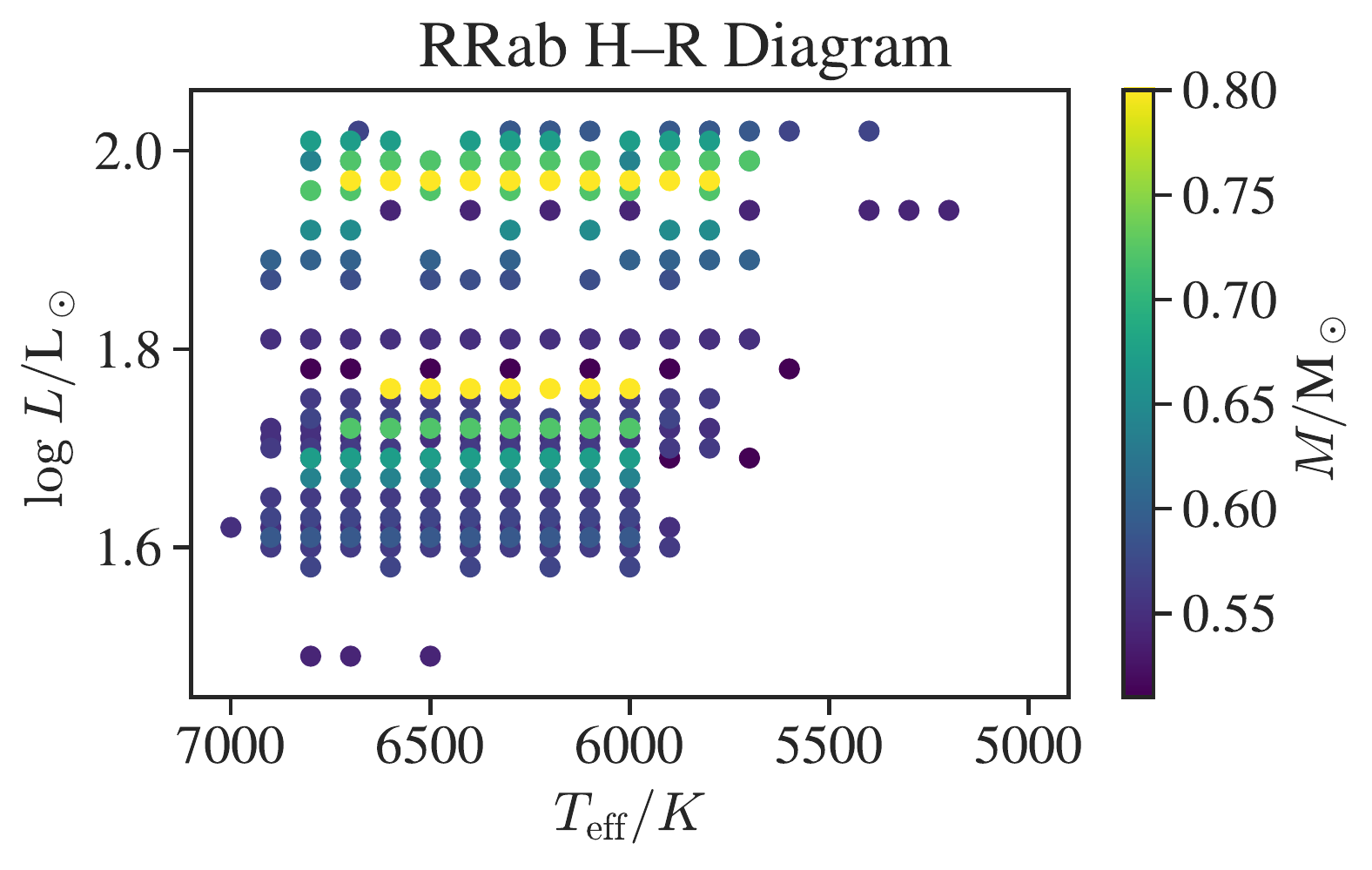}%
    \caption{Hertzsprung--Russell diagrams for the grids of models. Each point is a stellar model whose light curve properties have been computed.}
    \label{fig:HR}
\end{figure}

\begin{figure*}
    \centering
    \includegraphics[width=0.5\linewidth,keepaspectratio]{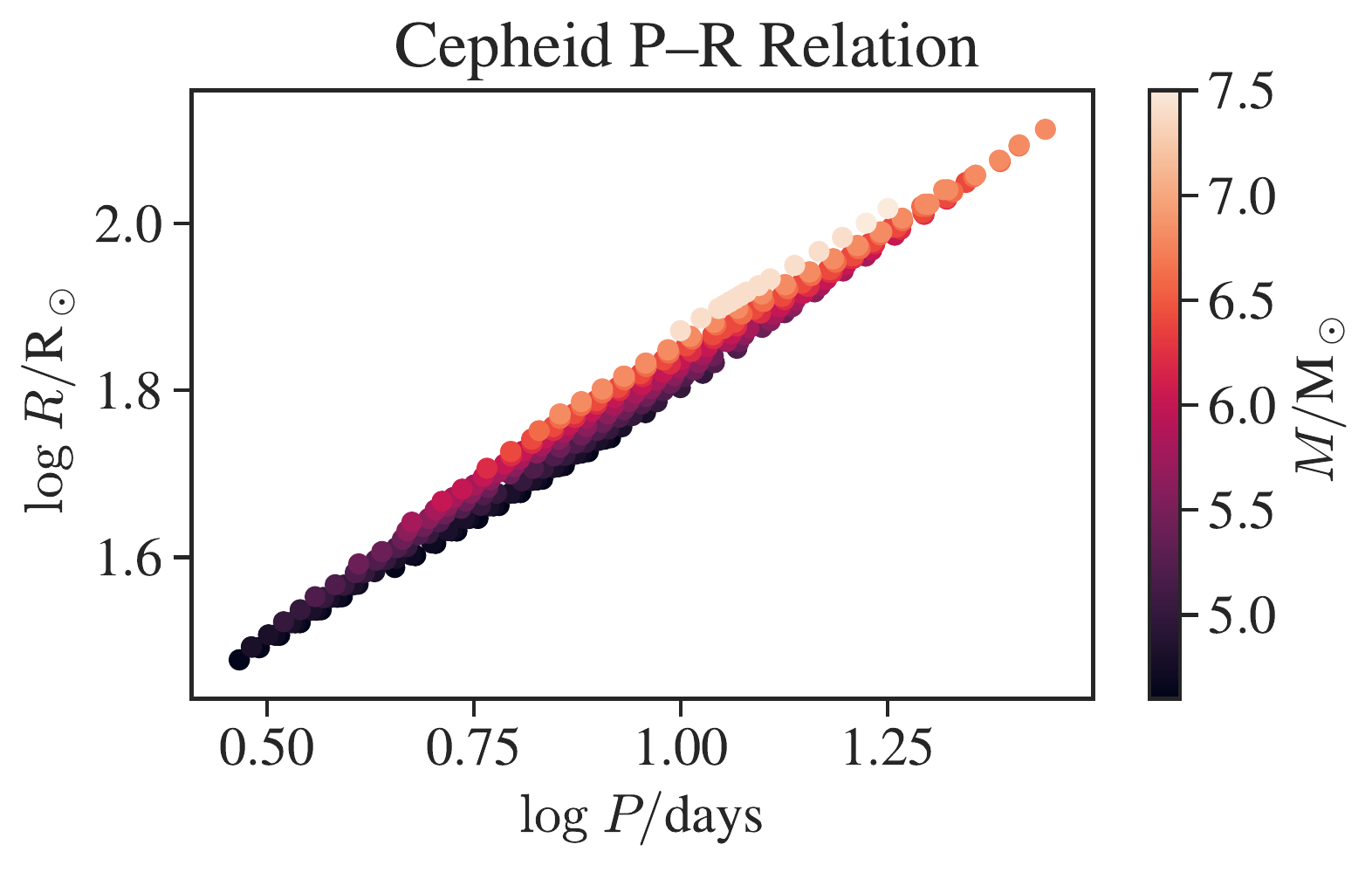}%
    \includegraphics[width=0.5\linewidth,keepaspectratio]{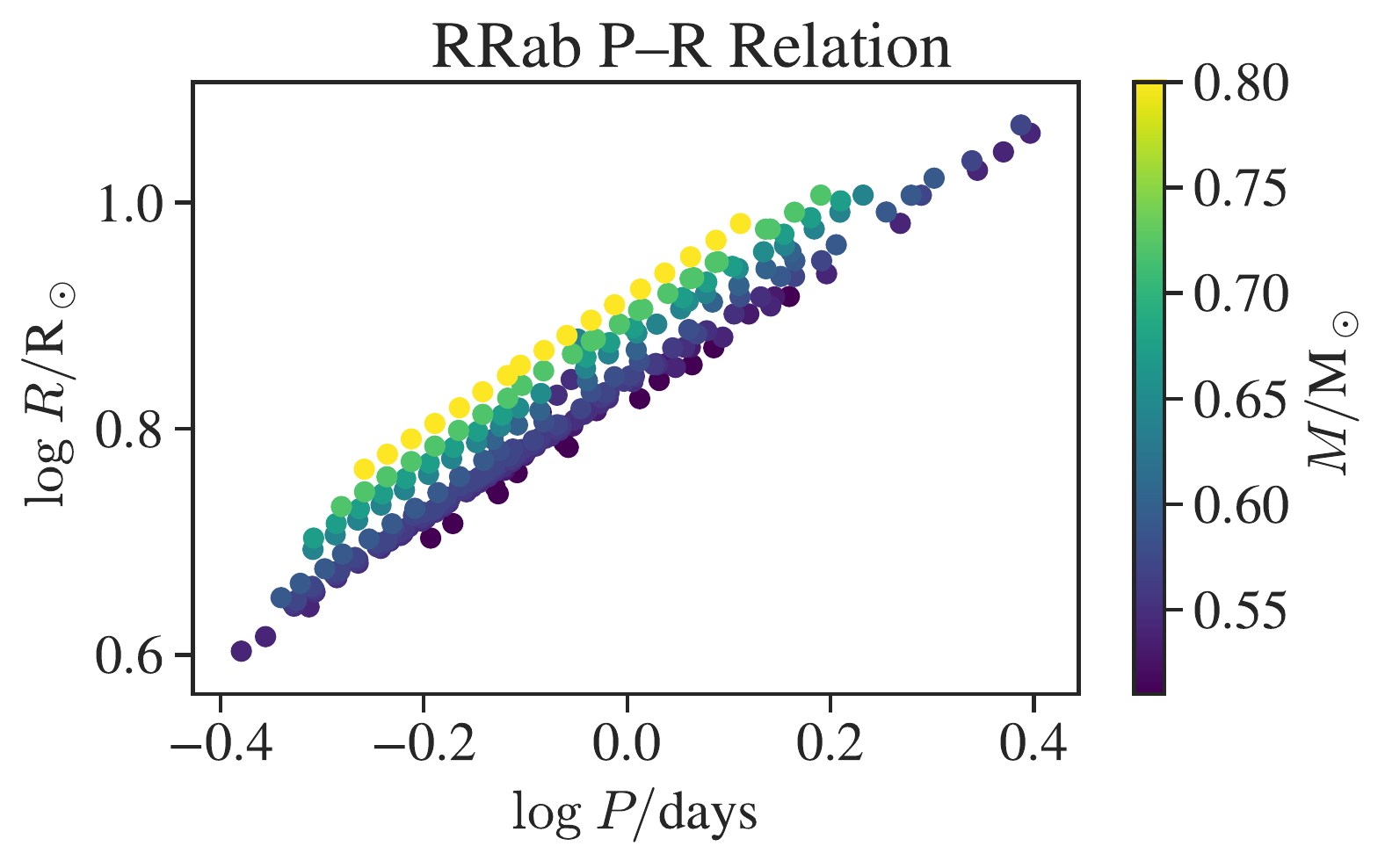}\\%
    \includegraphics[width=0.5\linewidth,keepaspectratio]{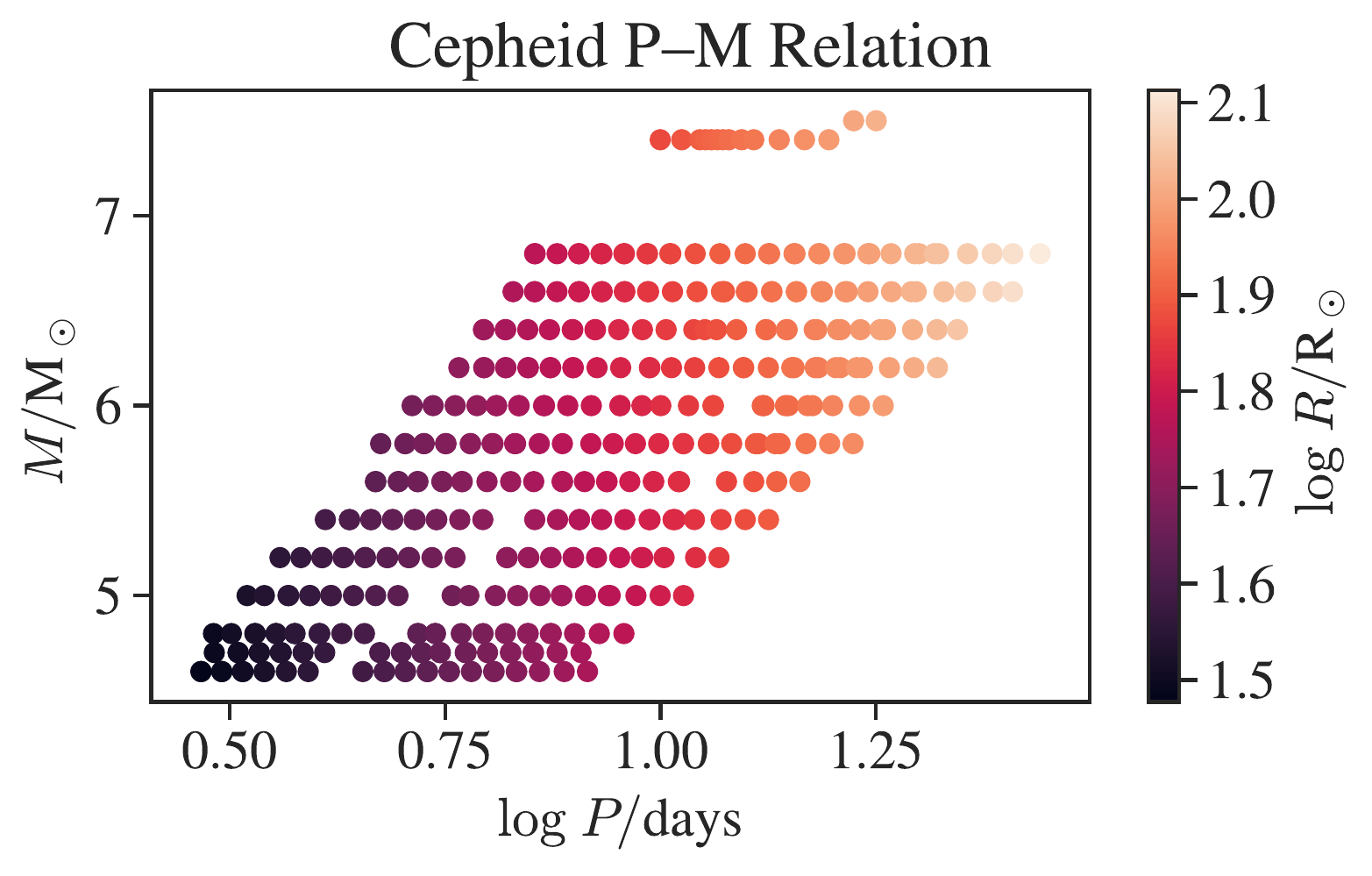}%
    \includegraphics[width=0.5\linewidth,keepaspectratio]{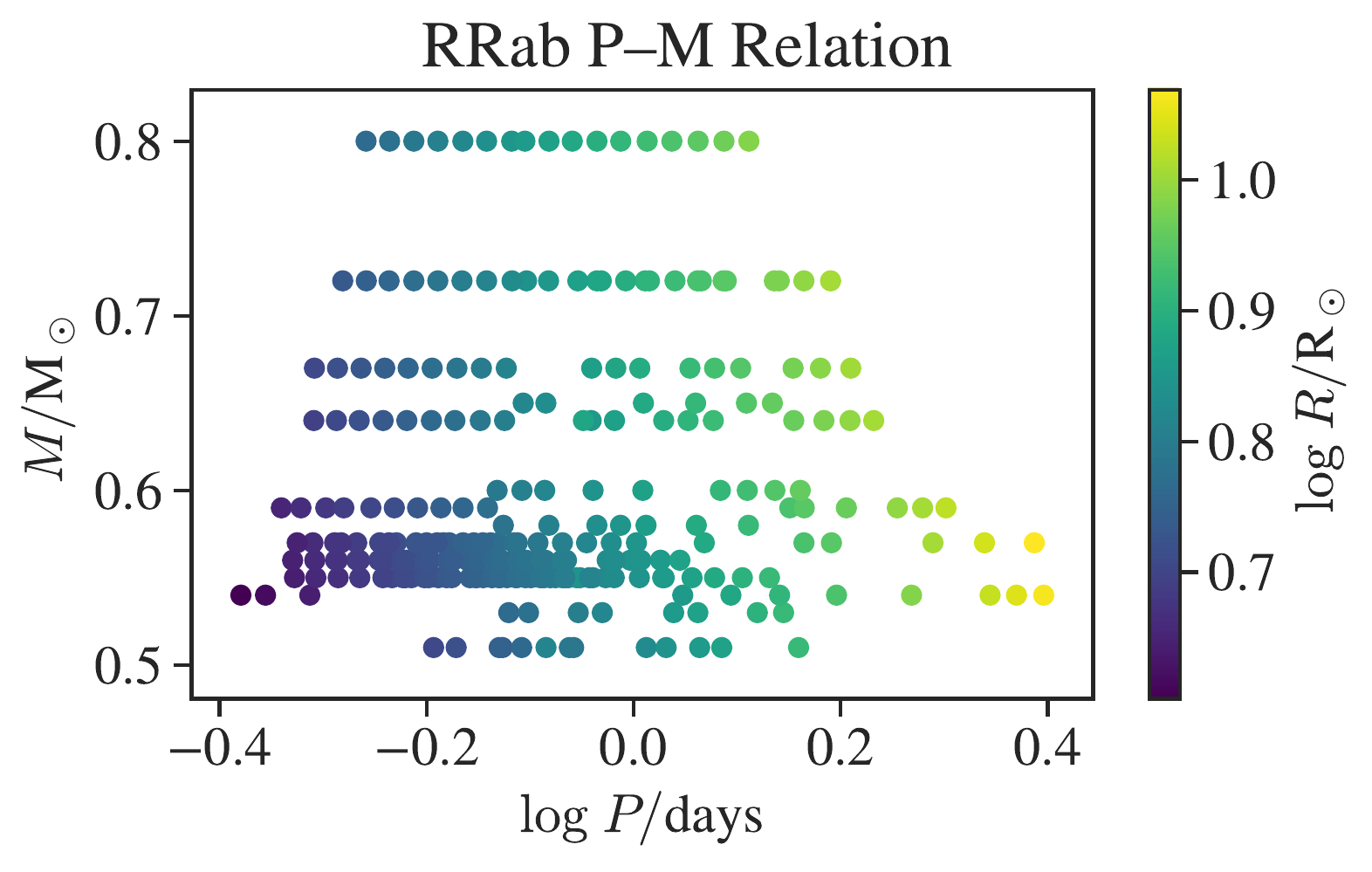}%
    \caption{Theoretical period--radius (top) and period--mass (bottom) diagrams for Cepheid (left) and RR~Lyrae (right) models. The radius is strongly constrained by the period, but with some scatter due to variations in the stellar mass. The mass, on the other hand, is only a weak function of the period, with much scatter due to variations in the stellar radius. \label{fig:PRM}}
\end{figure*}

Figure~\ref{fig:HR} displays the input ``data'': an HR diagram of the two grids of models (one for Cepheids and RR~Lyraes respectively). 
Figure~\ref{fig:PRM} presents the period--radius and period--mass relations displayed by the models. 
The radius is strongly constrained by the period, with some scatter (especially in the RR~Lyrae case) due to mass. 
These relations are essentially consequences of the period--mean density relation.

The observational data used in this study are again described in \citet{2017MNRAS.466.2805B} and \citet{2018MNRAS.481.2000D}, and consist of $V$- and $I$-band data from the fourth phase of the Optical Gravitational Lensing Experiment \citep[OGLE-IV,][]{2014AcA....64..177S, 2015AcA....65..297S, 2016AcA....66..131S} for fundamental mode Cepheids in the LMC and RR~Lyraes in the LMC, SMC, and Galactic bulge. A Fourier decomposition according to Equation~\ref{eq:Fourier} was applied to the data to obtain the Fourier parameters ($A_k, {\phi}_k$). 
Furthermore, the skewness and acuteness of each light curve was computed. 
The adopted periods were taken from the OGLE catalog.

\section{Methodology} \label{sec:methods}
Our methodology is to use modern methods from information theory, computer science, and artificial intelligence to look for relations connecting the period and measures of light curve structure to fundamental stellar parameters based on theoretical models. 
This approach follows similar recent works on the asteroseismology of solar-like stars \citep{2016ApJ...830...31B, 2019A&A...622A.130B, 2016MNRAS.461.4206V, 2017ApJ...839..116A}. 
After discussing the statistical significance of these relations and, in particular, the merit of light curve structure in estimating fundamental parameters, we apply these relations, developed from purely theoretical data, to observations from OGLE-IV. 

The grid of theoretical models have been computed via a forward function $f$ defined as
\begin{equation}
    f(M, L, R, T_{\text{eff}}, \ldots) = [\text{period},\; \text{light curve structure}]
\end{equation}
with $f$ representing the equations of stellar pulsation radiation hydrodynamics. 
The light curve structure contains the $I$- and $V$-band amplitudes, acutenesses, skewnesses, and the coefficients $A_1$, $A_2$, and $A_3$. 

We now wish to invert this relation. 
In particular, we seek the function $g=f^{-1}$, i.e.,
\begin{equation}
    g(\text{period},\; \text{light curve structure}) = [M, L, R, T_{\text{eff}}, \ldots] 
\end{equation}
which is not guaranteed to exist. 
To approximate this function we use an artificial neural network (ANN) \mb{trained using scikit-learn \citep{scikit-learn}} on the grid of theoretical models \citep[for an overview of ANNs, see, e.g.,][]{hastie2005elements}. 
We employ two hidden layers, with each having 100 hidden neurons containing rectified linear unit \citep[ReLU,][]{nair2010rectified} activation functions. 
We train the ANN using the limited-memory Broyden--Fletcher--Goldfarb--Shanno algorithm \citep[L-BFGS,][]{Liu1989} until the squared loss reached a tolerance of $10^{-7}$. 
\mbb{We use $L_2$ regularization to penalize large network weights, which reduces the impact of non-relevant attributes. 
We use a regularization tuning parameter of $0.0001$, but we note that different choices for this parameter make little impact on the results.} 
As ANNs are sensitive to data scalings, we preprocess every variable by subtracting the median and dividing by the median absolute deviation.

\subsection{\mbb{Model Assessment}}
We now seek to evaluate how well the ANN can predict physical parameters based on the theoretical models and using the theoretical period and light curve structure. 
As a baseline, we will compare with a linear model (LM) based on \mb{the logarithm of the} period, e.g., 
\begin{equation}
    y = a + b \log P
\end{equation}
where $y$ is a quantity we wish to estimate (e.g., radius) and $a$ and $b$ are the coefficients of the fit. 
\mbb{We will evaluate these estimator models using two-fold cross validation. 
This method works by fitting the estimator using half of the theoretical models, and then subsequently estimating the parameters of the left-out models. 
The procedure is then repeated by swapping the training set with the testing set. 
Cross validation helps to assess whether the ANN is over-fitting the training data, as in that case it would produce poor assessments on the test data. }
We can then quantify the standard deviation of the errors as well as the coefficient of variation 
\begin{equation}
    R^2 = 1 - \frac{\sum_i (y_i - \hat y_i)^2}{\sum_i (y_i - \bar y)^2}
\end{equation}
where $\hat y$ is the estimated value of $y$ of the left-out models (made either by the LM or by the ANN), and $\bar y$ is the mean value of $y$. 
Note that $R^2$ has a maximum value of one and is unbounded from below. 
An $R^2$ of zero implies that the estimator does no better than guessing the mean value of the variable being estimated.

\begin{figure*}
    \centering
    \includegraphics[width=0.5\linewidth]{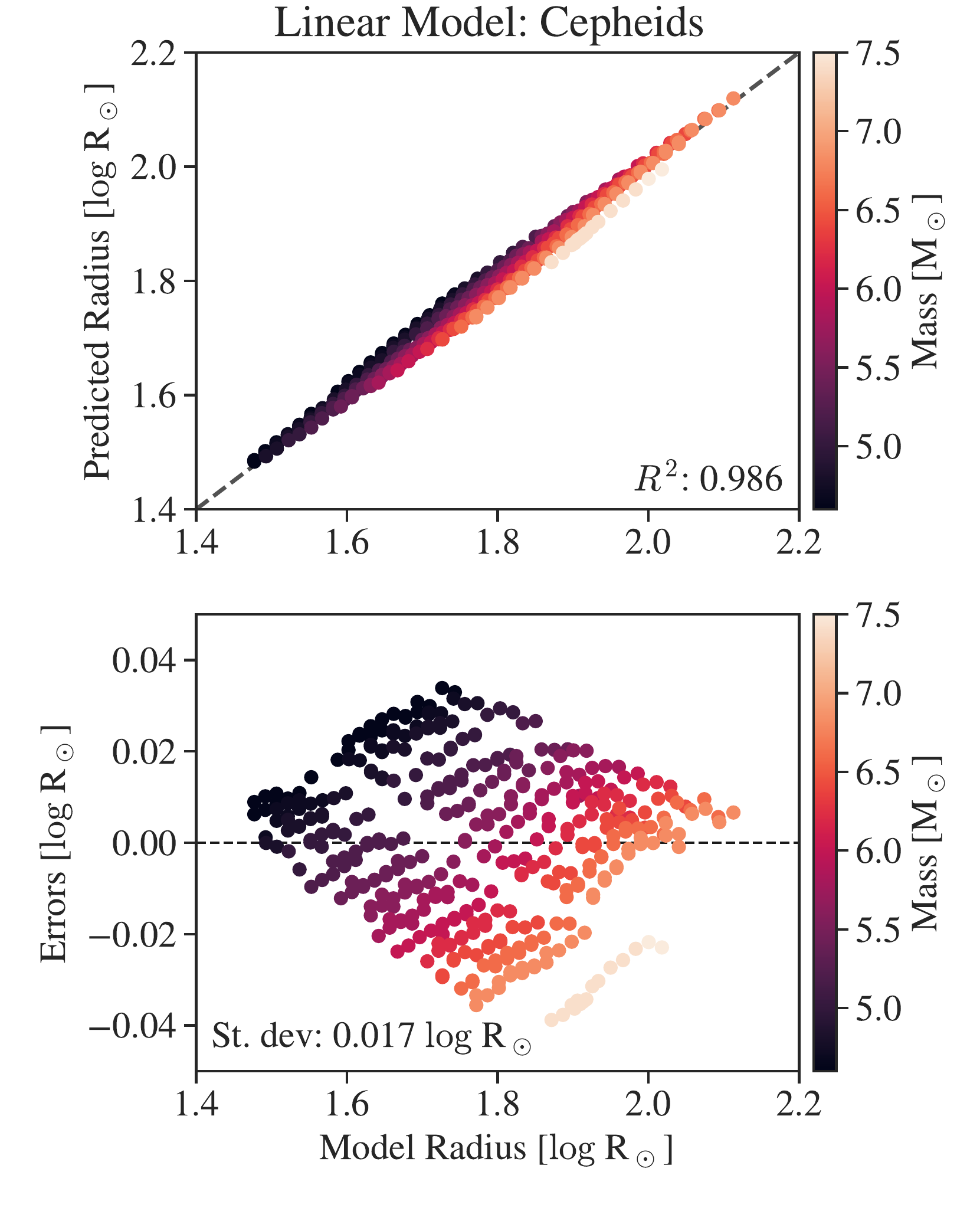}%
    \includegraphics[width=0.5\linewidth]{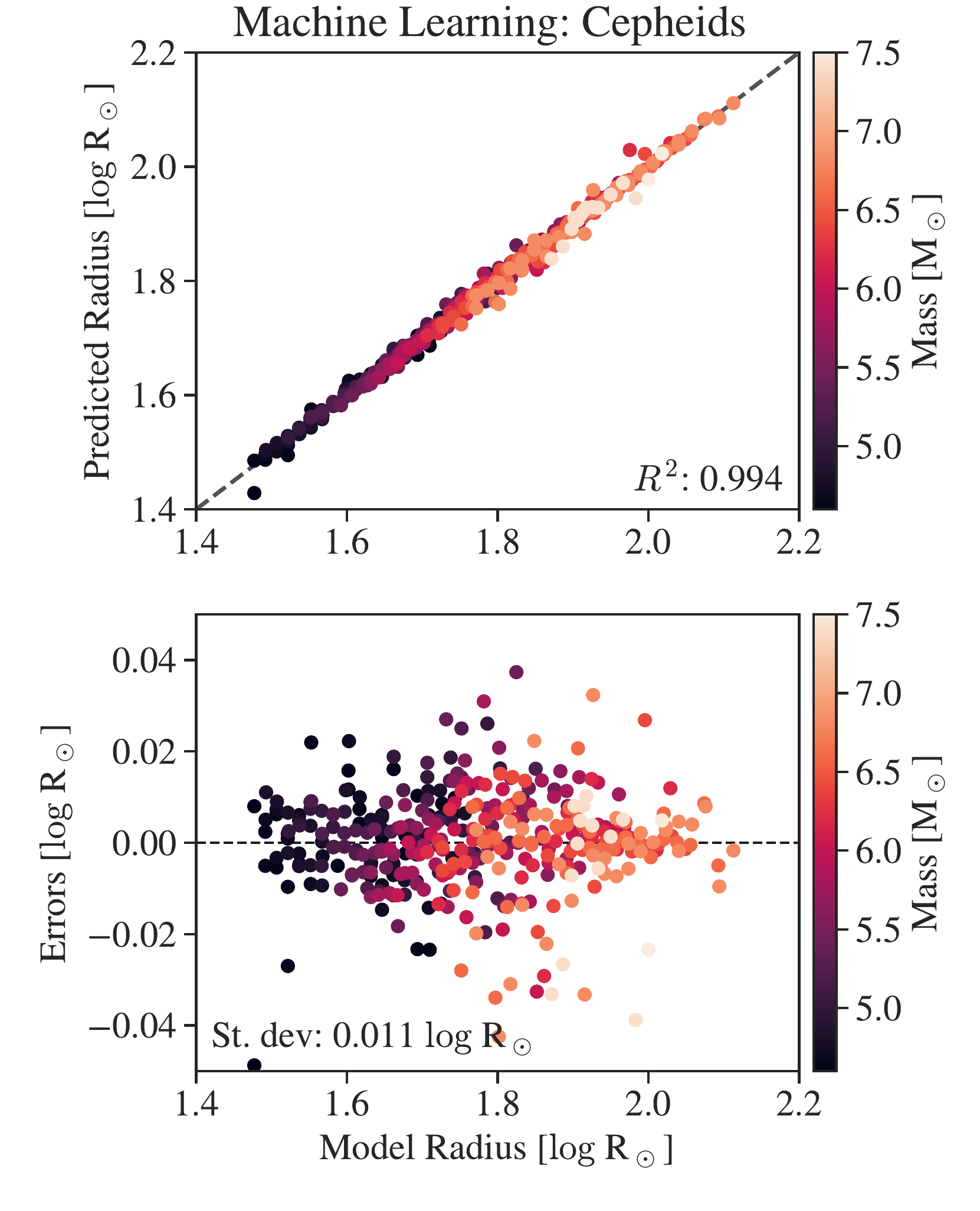}%
    \caption{Predicted versus actual radius for fits based on only the period (left panels) and fits using machine learning based on the period and the light curve structure (right panels). The bottom panels show the (out-of-bag) errors of the relations. \label{fig:CEP-radius}}
    \vspace*{\baselineskip}
    \centering
    \includegraphics[width=0.5\linewidth]{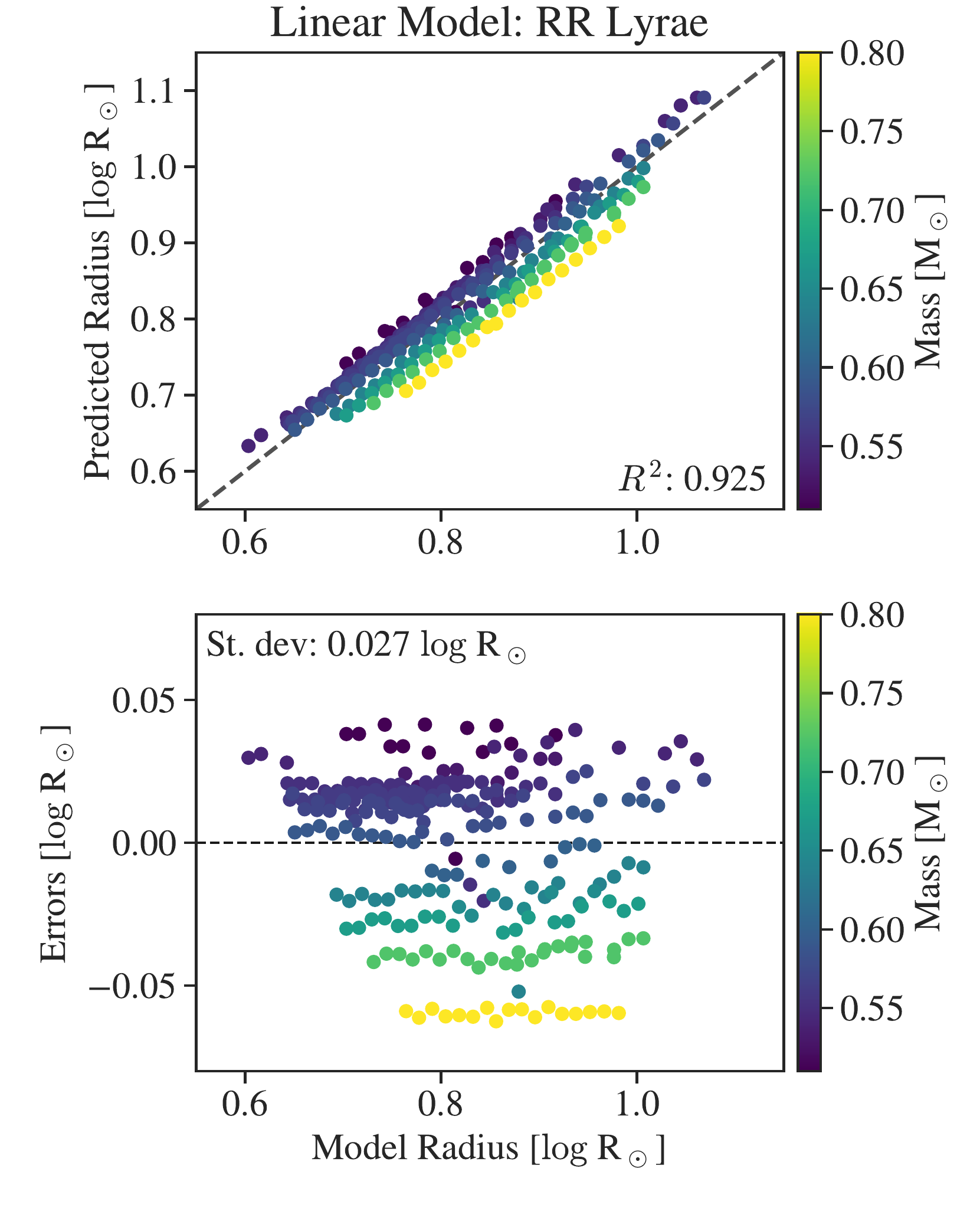}%
    \includegraphics[width=0.5\linewidth]{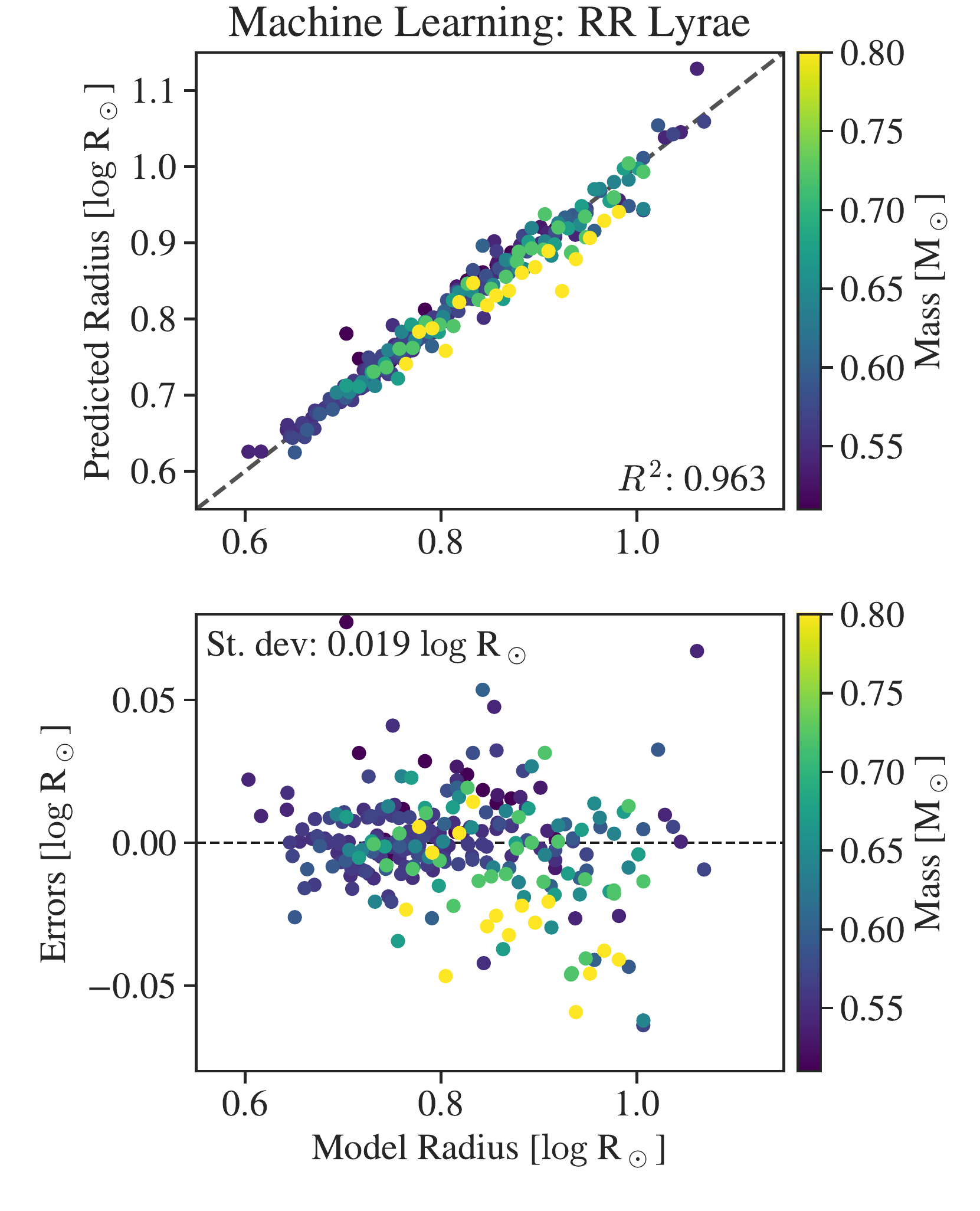}%
    \caption{The same as Figure~\ref{fig:CEP-radius} but for RR~Lyrae models. \label{fig:RRL-radius}}
\end{figure*}

The left panels of Figures~\ref{fig:CEP-radius} and \ref{fig:RRL-radius} display radii estimated using cross validation with an LM based only on the period. 
The right panels show a machine learning model which uses period and features of the light curve structure to estimate radii. 
Whilst we see a modest increase in $R^2$ with the addition of light curve structure, we see that the structure in the errors evident in the bottom left panel (predicting radius using only period) is no longer present in the bottom right panel (using period and light curve structure). 
Note that in the bottom left panel, the dots of constant color are models with the same mass--luminosity pair but a different temperature. 

\begin{figure*}
    \centering
    \includegraphics[width=0.5\linewidth]{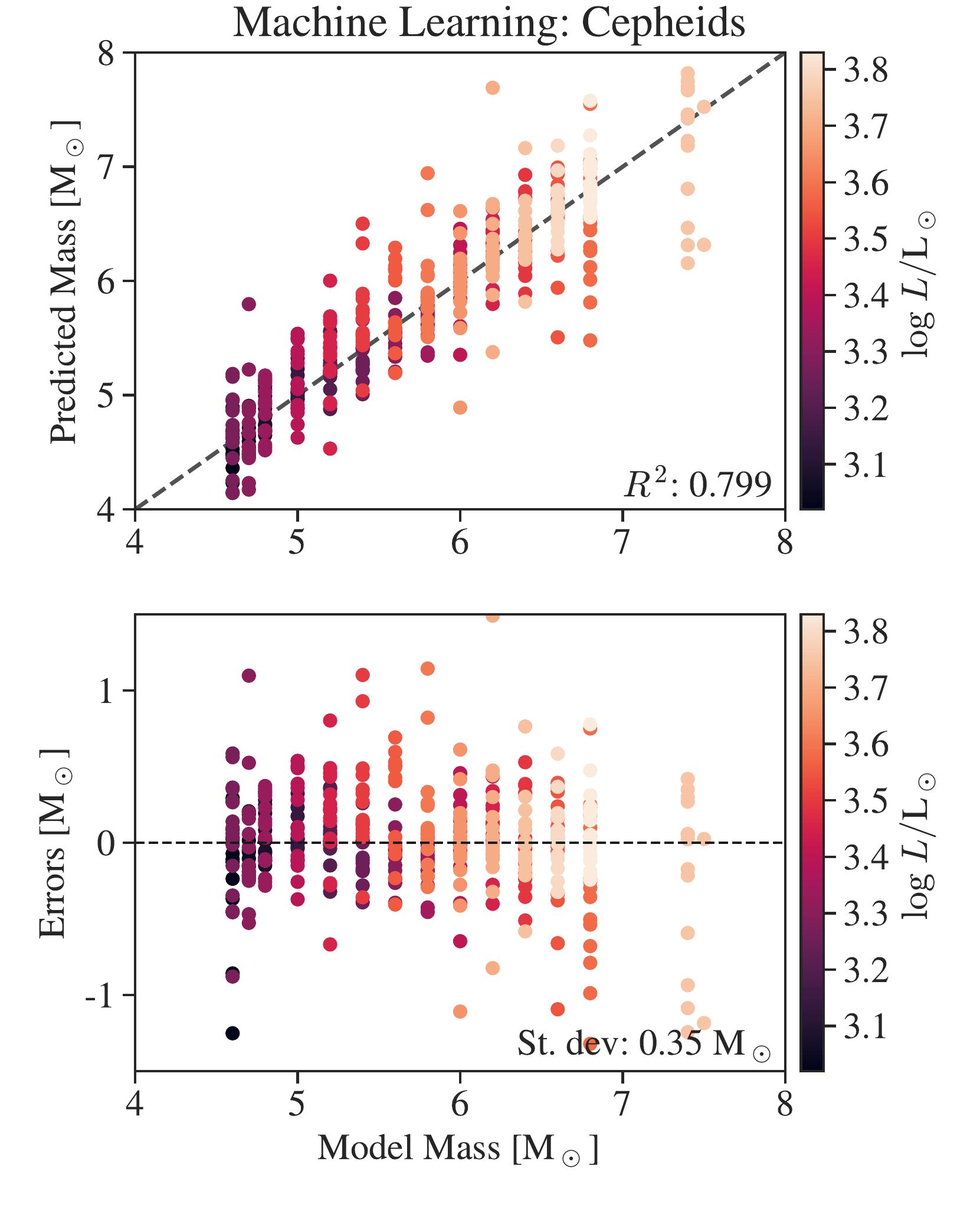}%
    \includegraphics[width=0.5\linewidth]{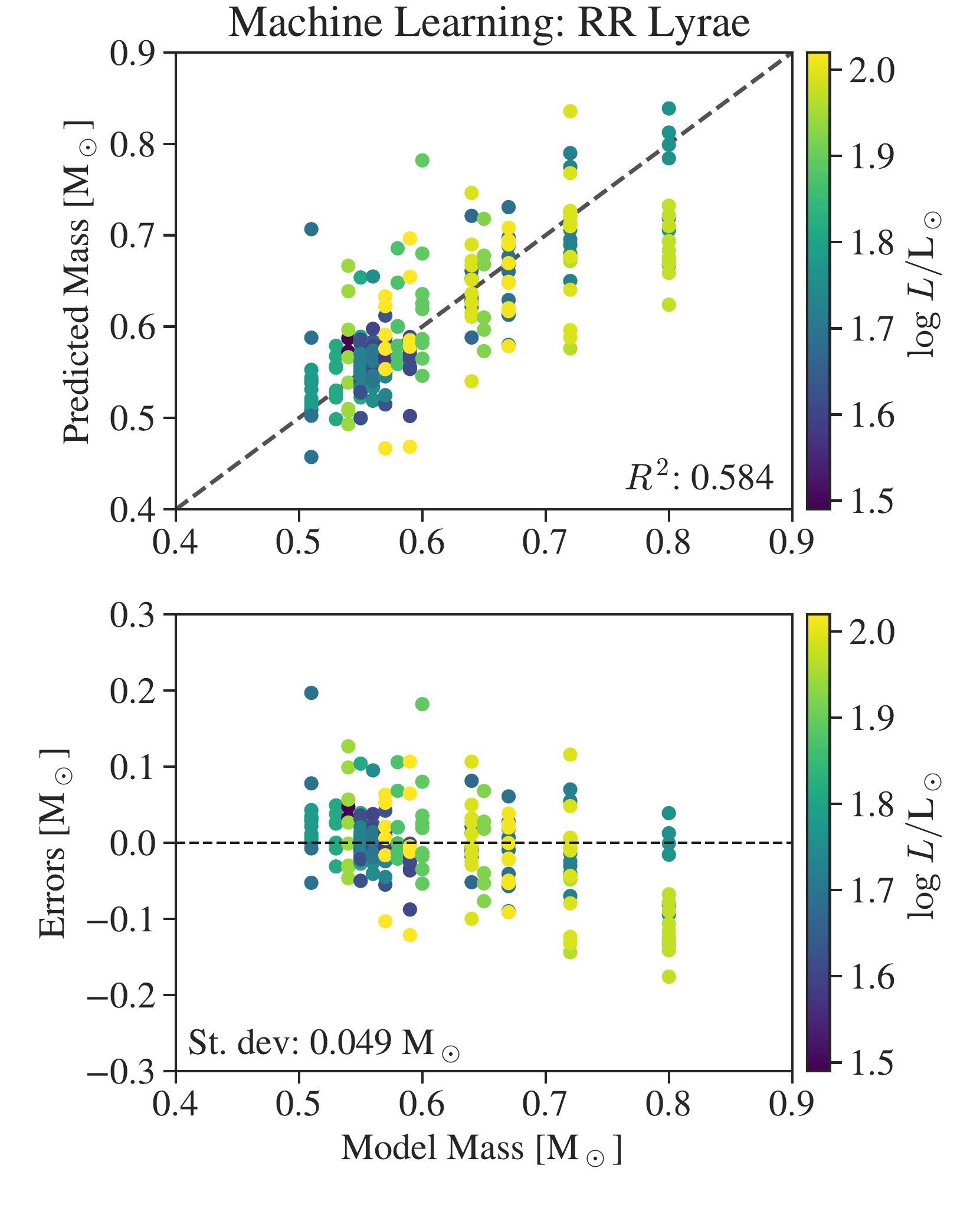}%
    \caption{Predicted versus actual mass for estimates based on the period and light curve structure using machine learning on theoretical models. }
    \label{fig:CEP-RRL-mass}
\end{figure*}

In Figure~\ref{fig:CEP-RRL-mass}, we present results related to predicting mass for both RR~Lyrae and Cepheid stars. 
In this case, we do not show the LM using just period as a dependent variable because these results would be very poor (\emph{cf}.~Figure~\ref{fig:PRM}). 
Again we see the destruction of structure in the errors, and find strong statistical significance when adding light curve structure to period for both Cepheids and RR~Lyraes.

\begin{figure*}
    \centering
    \includegraphics[width=0.5\linewidth]{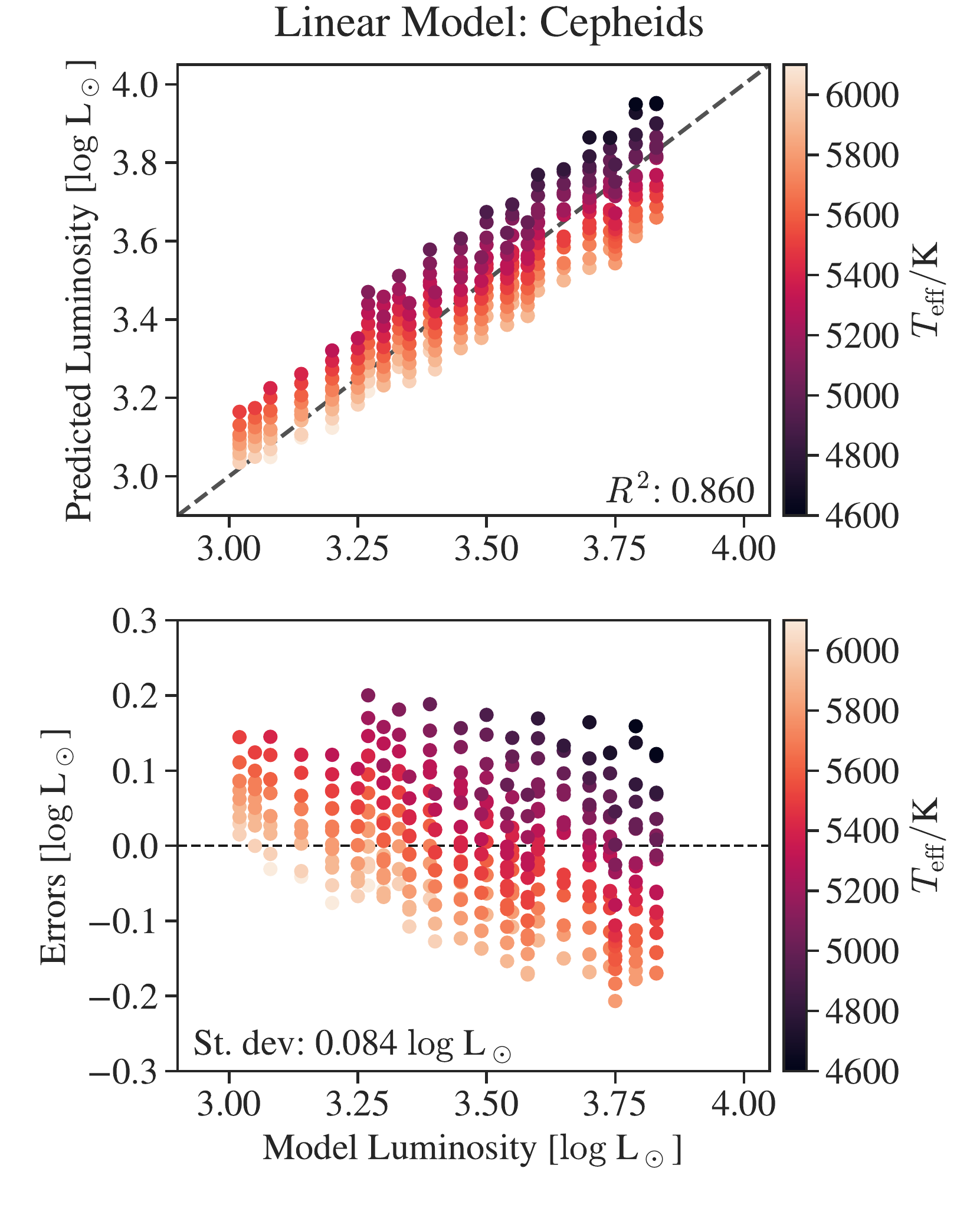}%
    \includegraphics[width=0.5\linewidth]{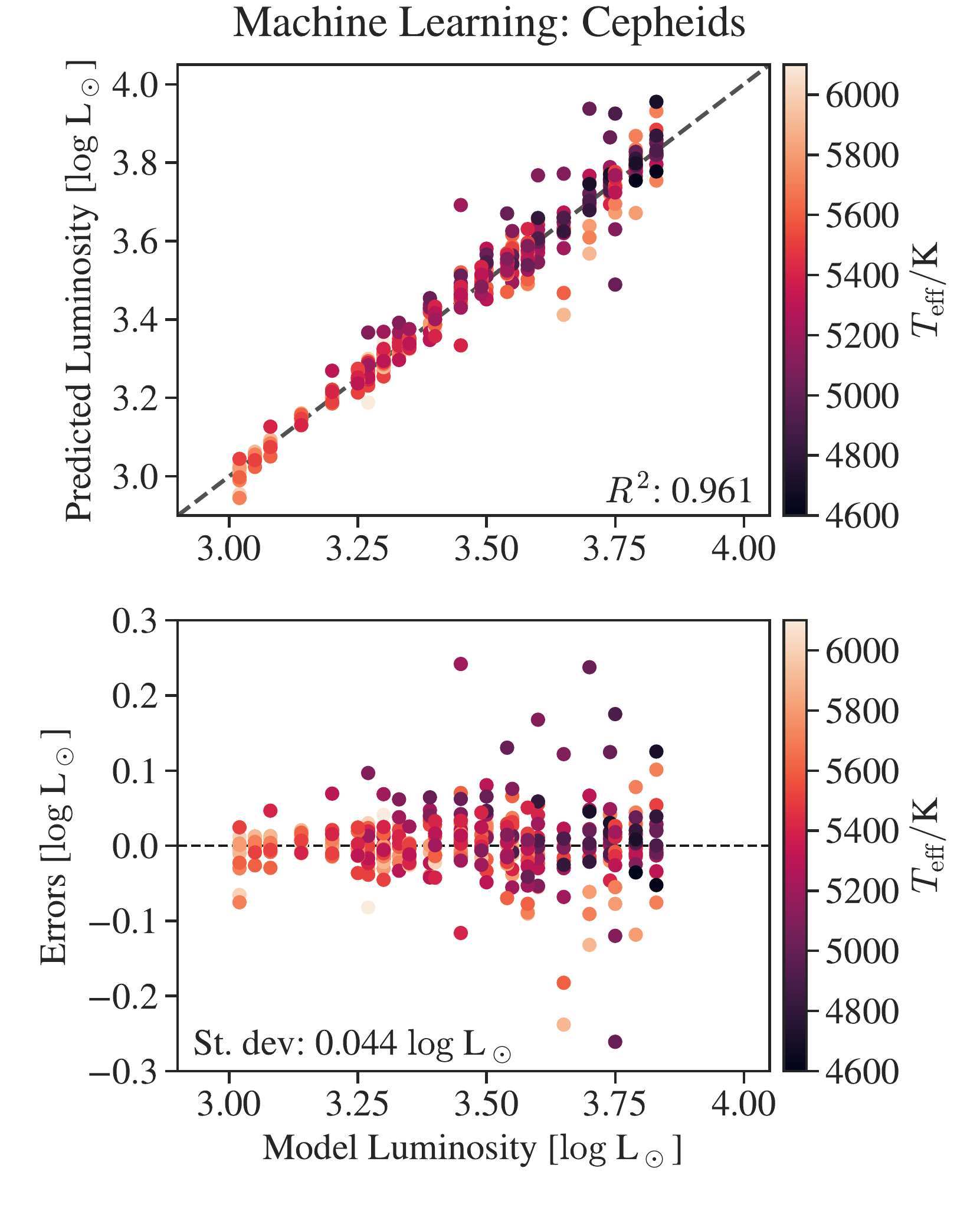}%
    \caption{Predicted versus actual luminosities for estimates made based on the PL relation (left) and period and light curve structure (right). The errors of the PL relation are scattered due to differences in the effective temperature, whereas those using light curve structure show no such trends. \label{fig:CEP-L}}
    \vspace*{\baselineskip}
    \includegraphics[width=0.5\linewidth]{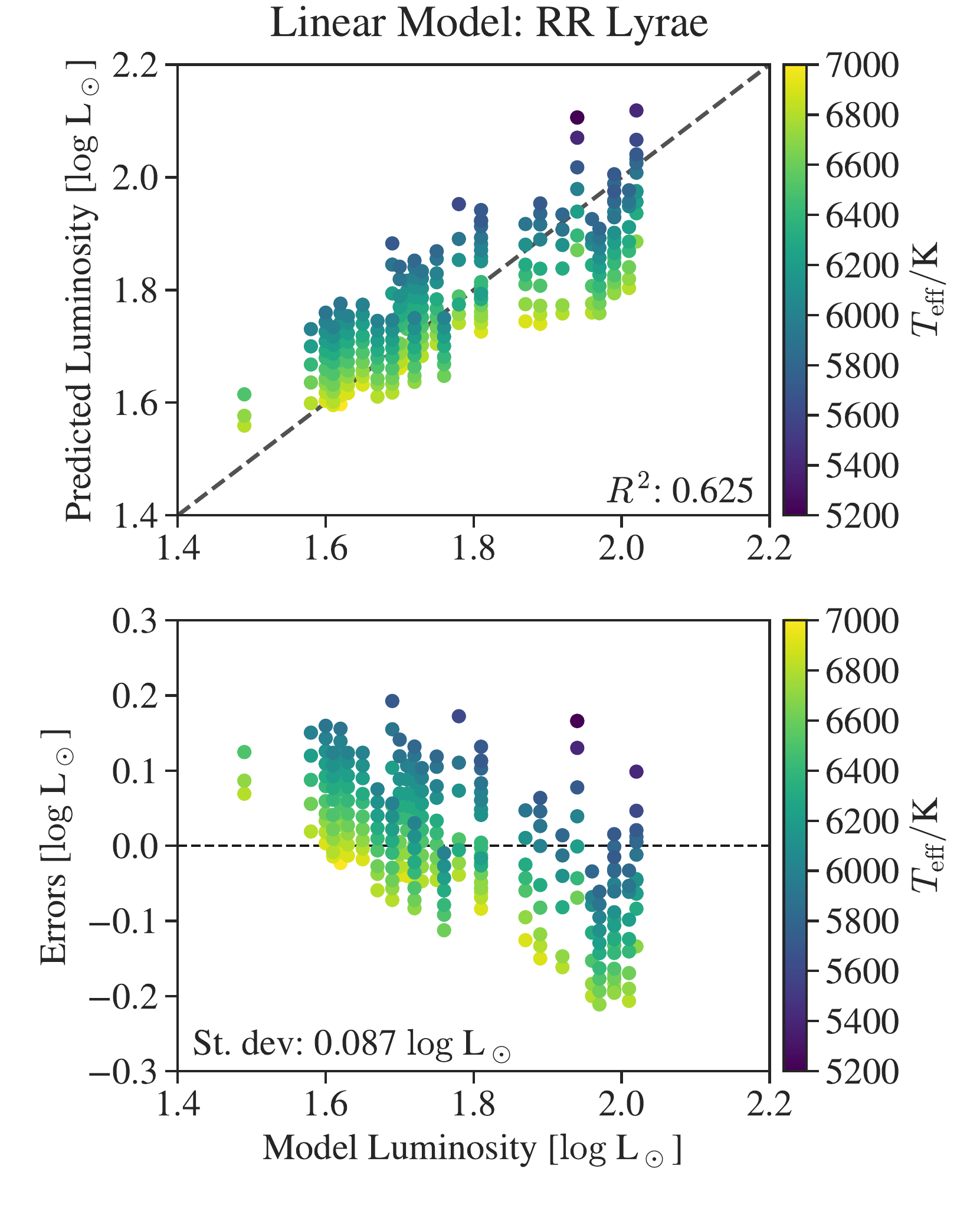}%
    \includegraphics[width=0.5\linewidth]{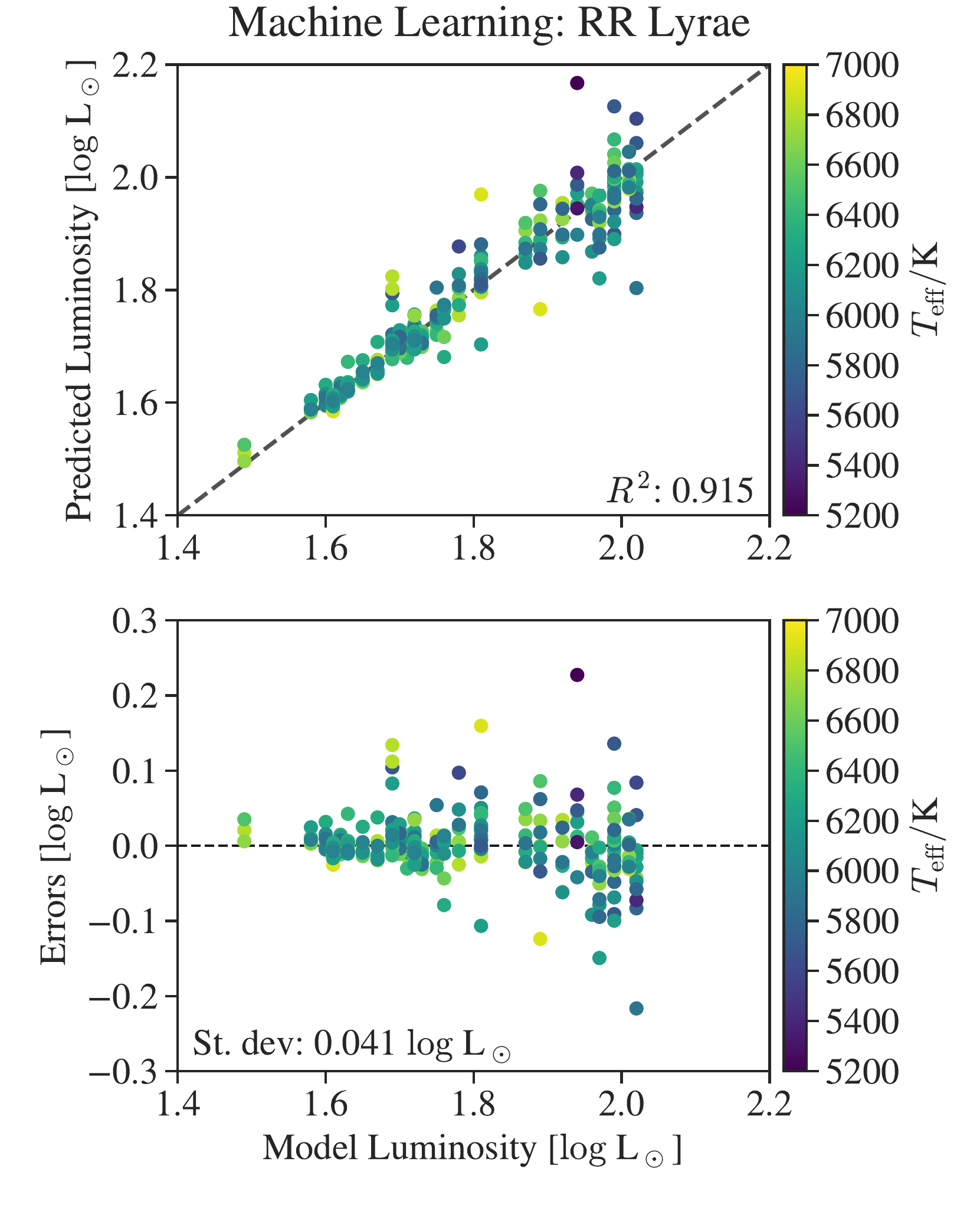}%
    \caption{The same as Figure~\ref{fig:CEP-L} but for RR~Lyrae models. \label{fig:RRL-L}}
\end{figure*}

\begin{figure*}
    \centering
    \includegraphics[width=0.5\linewidth]{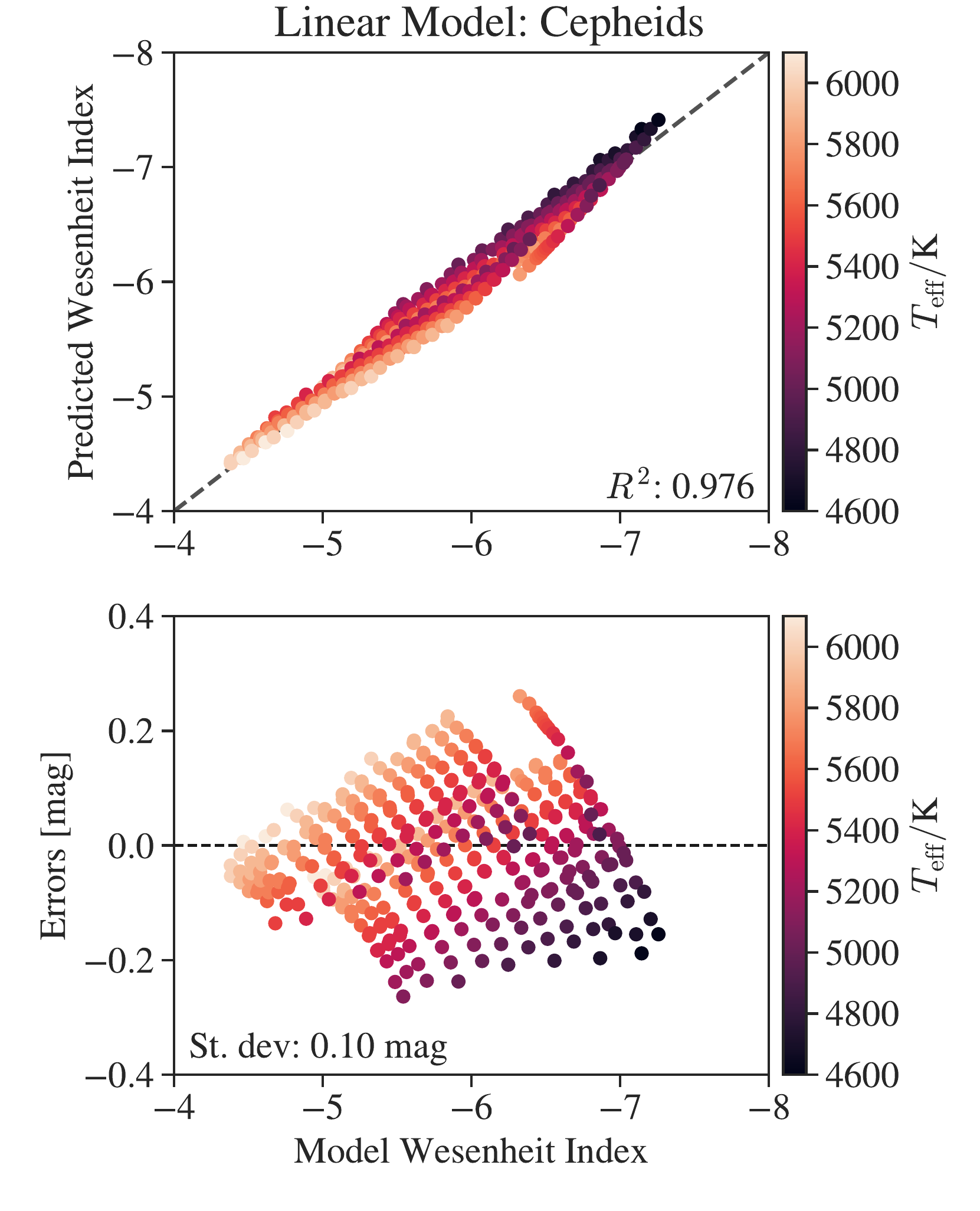}%
    \includegraphics[width=0.5\linewidth]{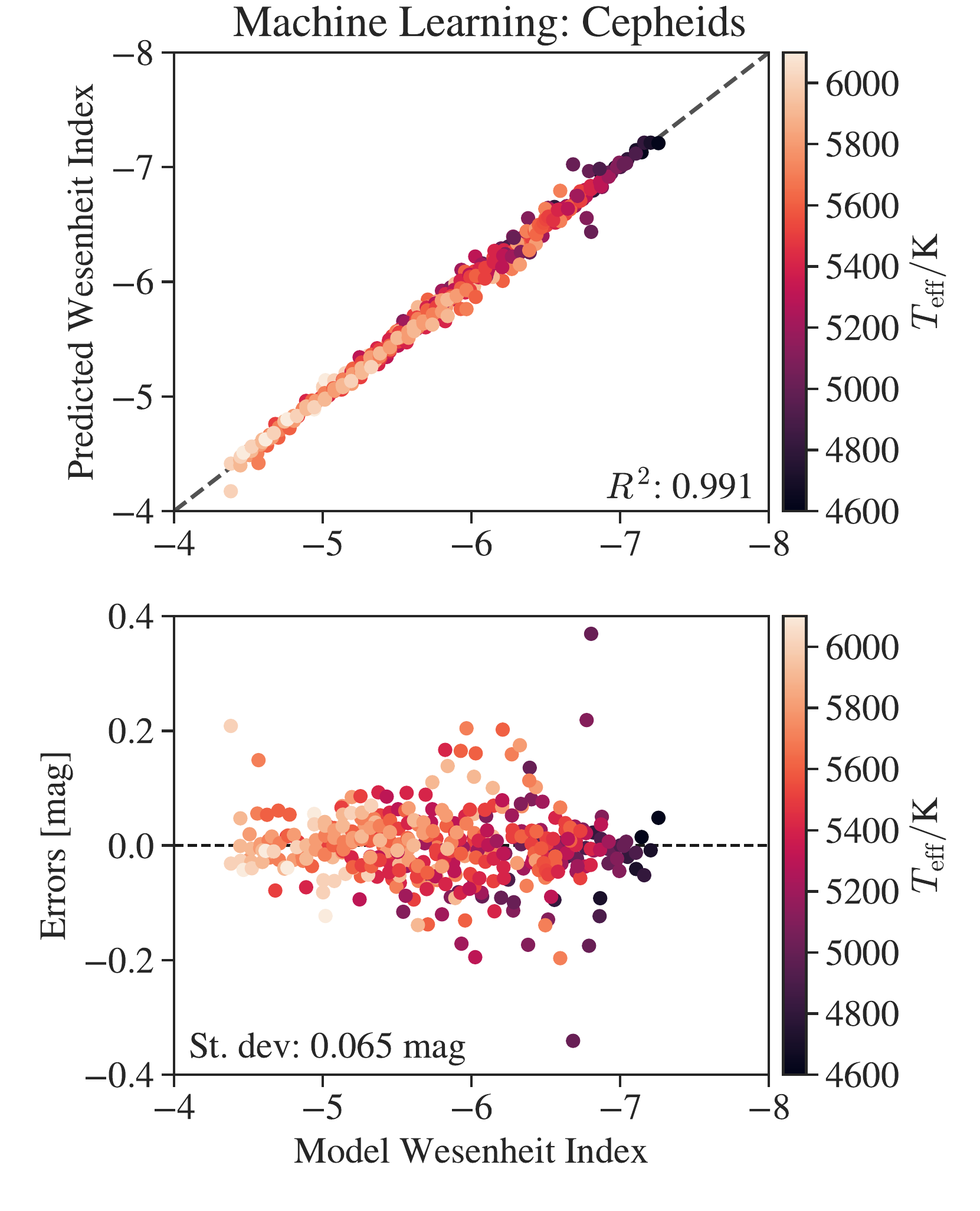}%
    \caption{Predicted versus actual Wesenheit indices for estimates made based on the period (left) and period and light curve structure (right). \label{fig:CEP-W}}
    \vspace*{\baselineskip}
    \includegraphics[width=0.5\linewidth]{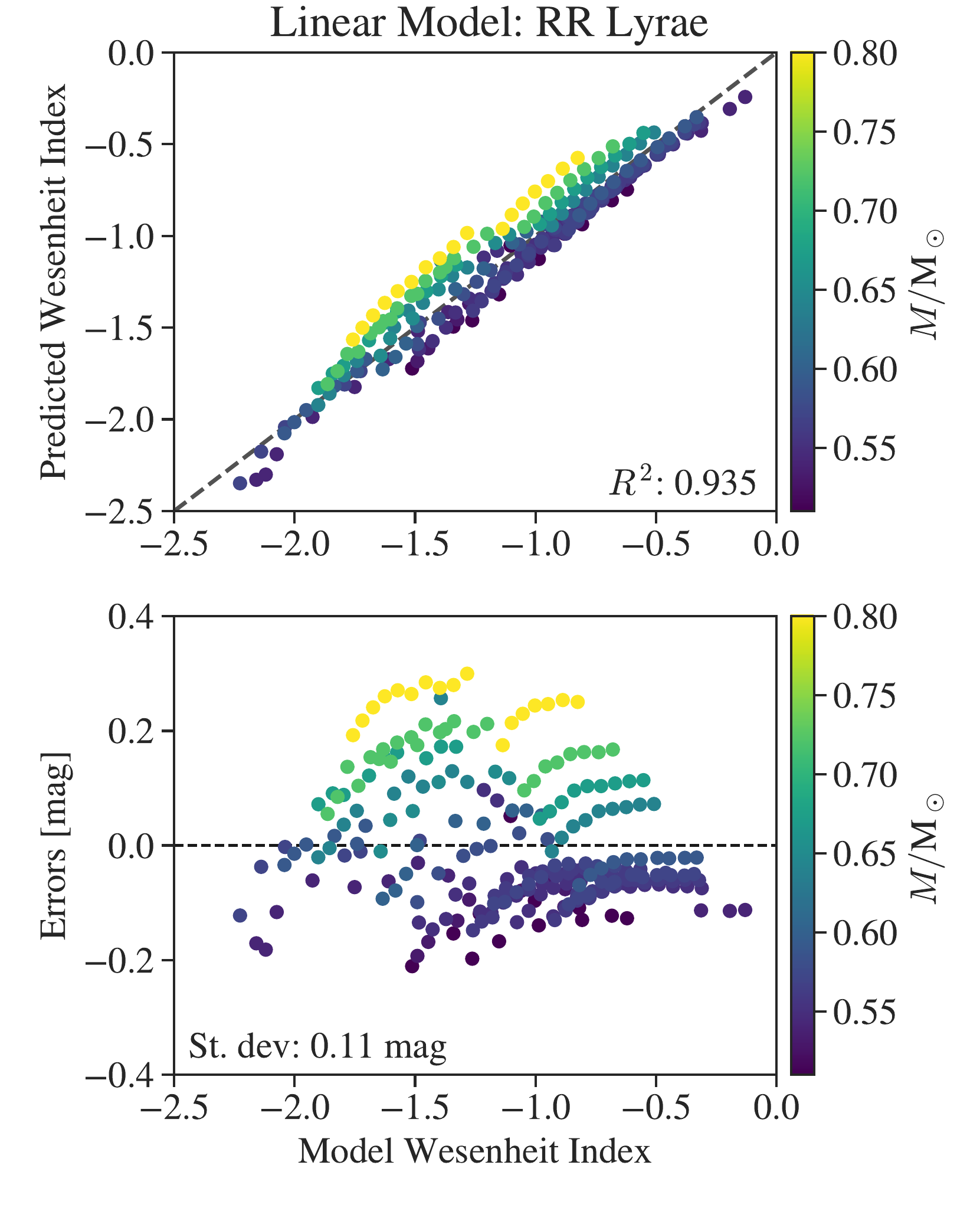}%
    \includegraphics[width=0.5\linewidth]{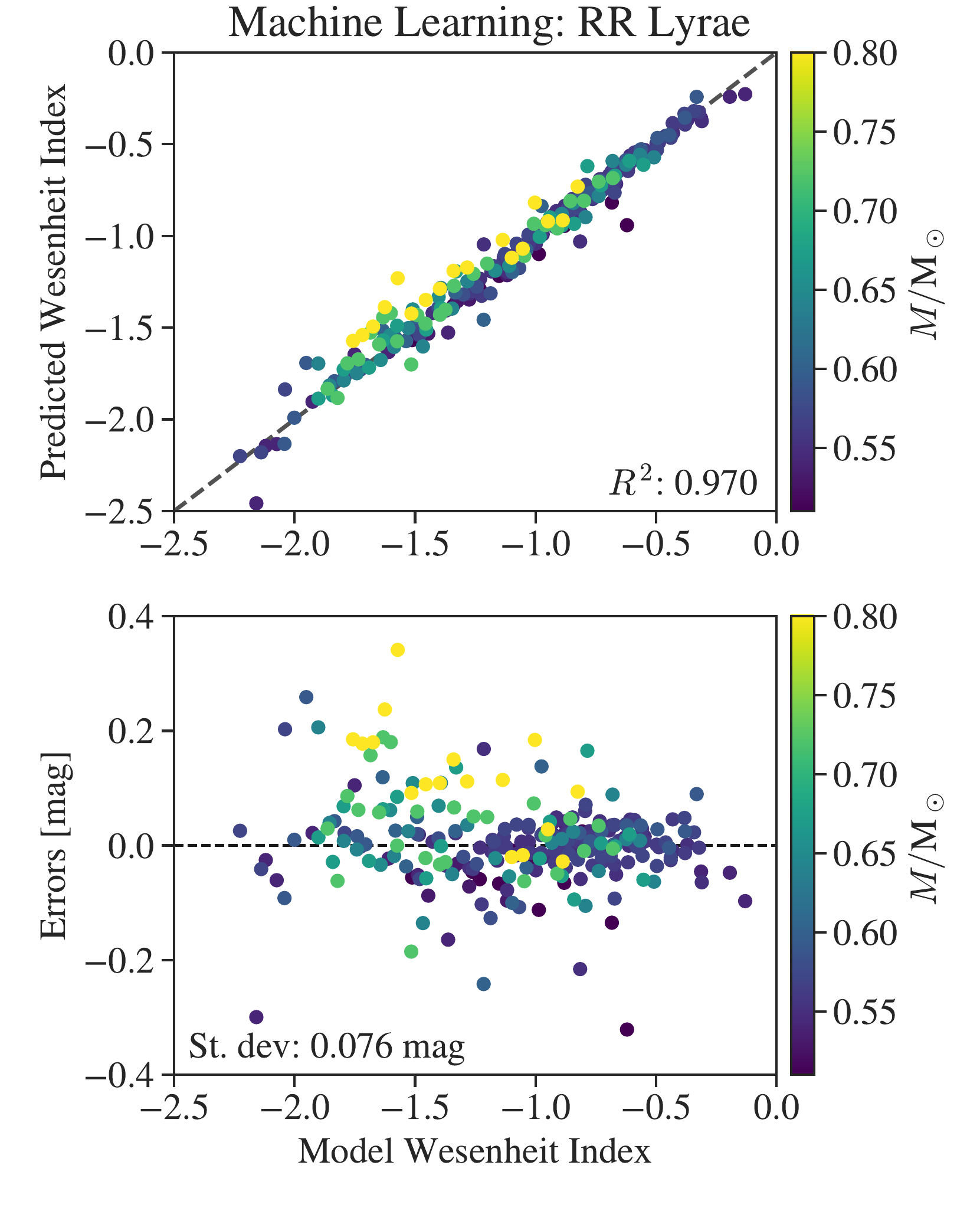}%
    \caption{The same as Figure~\ref{fig:CEP-W} but for RR~Lyrae models. \label{fig:RRL-W} }
\end{figure*}

In the top panels of Figure~\ref{fig:CEP-L} we present a comparison of predicted and actual theoretical luminosities using just period and also using period and light curve structure. The errors associated with these predictions are given in the bottom panels. When we incorporate light curve structure, we see a large increase in $R^2$ and the reduction of structure in the errors. Figure~\ref{fig:RRL-L} presents similar results for RR~Lyraes. 
Similarly, Figures~\ref{fig:CEP-W} and ~\ref{fig:RRL-W} show important improvements when predicting the reddening-independent Wesenheit index \citep{1982ApJ...253..575M}
\begin{equation} \label{eq:W}
    W = I - 1.55\cdot (V-I)
\end{equation}
which is widely used in estimating distances \citep{2018AcA....68...89S}. 

\begin{figure*}
    \centering
    \includegraphics[width=\linewidth]{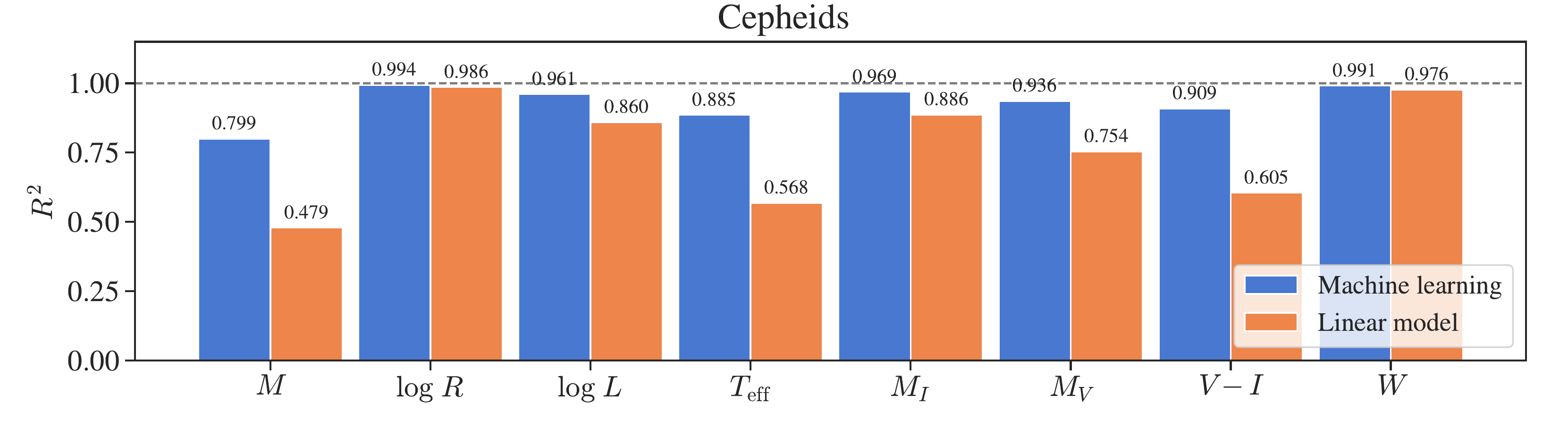}\\%
    \includegraphics[width=\linewidth]{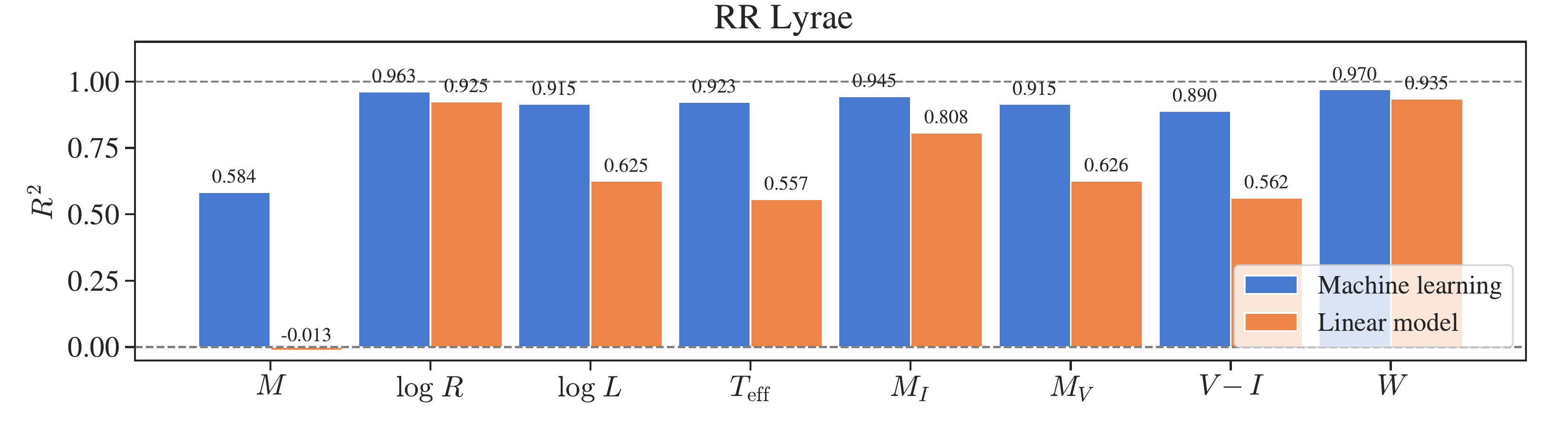}%
    \caption{Cross-validated coefficients of variation for fundamental parameters of Cepheids (top) and RR~Lyrae (bottom) comparing estimates of mass, radius, luminosity, effective temperature, I-band magnitude, V-band magnitude, color, and Wesenheit index made using an LM based only on period (orange) and using machine learning based on period and light curve structure (blue). Higher is better. In all cases, the use of information contained in the structure of the light curve significantly improves the estimate (see also Table~\ref{tab:significance}). \label{fig:loo-r2}} 
    \vspace*{1.5\baselineskip}
    \centering
    \includegraphics[width=\linewidth]{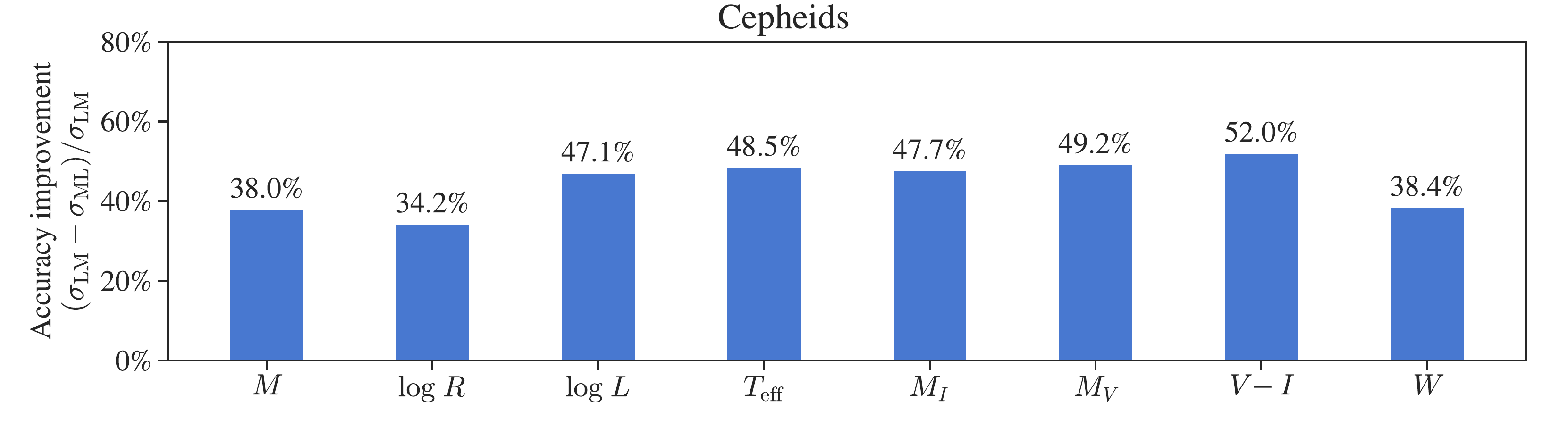}\\
    \includegraphics[width=\linewidth]{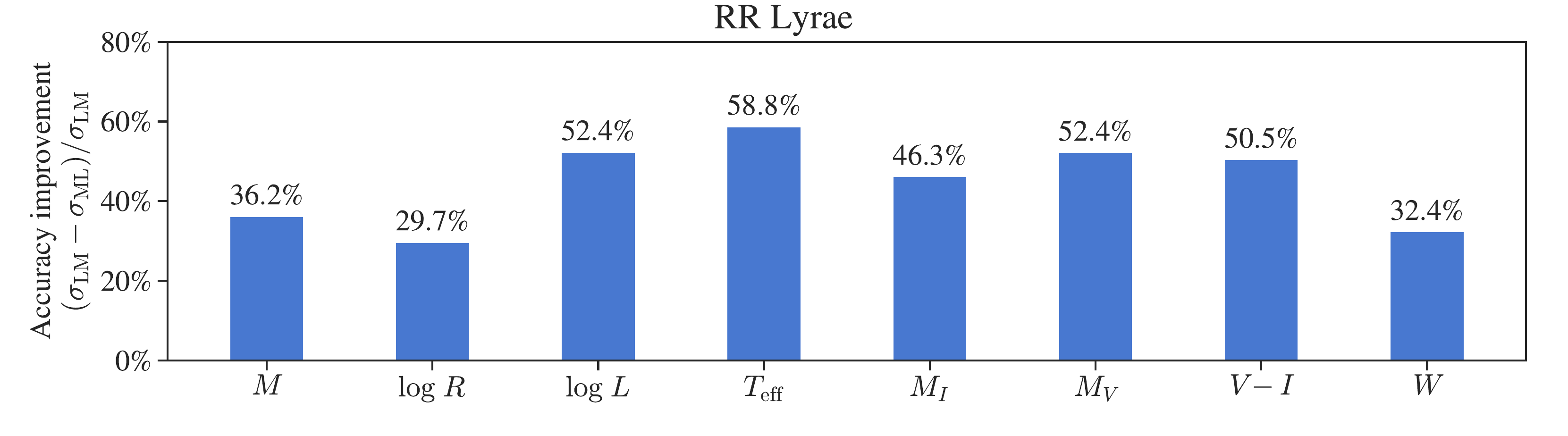}%
    \caption{Cross-validated improvements to the accuracy of estimates when information from the light curve structure is incorporated over an estimate made using only the period. Here $\sigma_{\text{LM}}$ refers to the standard deviation of the errors from an LM based on period, and $\sigma_{\text{ML}}$ refers to the standard deviation of the errors from a model based on period and light curve structure. \label{fig:loo-sds}} 
\end{figure*}

Figures~\ref{fig:loo-r2} and \ref{fig:loo-sds} summarize the findings over all of the fundamental parameters which we seek to estimate, which also include $I$- and $V$-band magnitudes and $(V-I)$ color. 
The former presents the improvements in $R^2$ when using period and light structure over just period when predicting various stellar parameters from the models.
The latter displays the improvements in the accuracy of our estimates when using period and light curve structure over just period when predicting various stellar parameters from the models.

\mbb{Another way of assessing the improvement of the ANN over the LM is to quantify what percentage of the estimates are improved. Table~\ref{tab:errors} lists these values, showing that in the great majority ($\sim 80\%$) of cases, the error of the ANN estimate is smaller than from the LM.} 

Finally, Table~\ref{tab:significance} table presents the standard deviations of the errors from cross validation. 
When estimating the parameters of an observed star, these standard deviations are to be added in quadrature with the random uncertainties arising from propagating the observational uncertainties through the ANN.

\begin{table}
    \centering
    \begin{tabular}{cll}
        Variable & CEP & RRL \\\hline\hline
        $M/$M$_\odot$ & 74\% & 81\%  \\
        $\log R/$R$_\odot$ & 74\% & 80\%  \\
        $\log L/$L$_\odot$ & 81\% & 80\%  \\
        $T_{\text{eff}}/$K & 85\% & 84\%  \\
        $M_I$ & 80\% & 81\%  \\
        $M_V$ & 82\% & 81\%  \\
        $V-I$ & 84\% & 82\%  \\
        $W$ & 76\% & 81\%  \\\hline
    \end{tabular}
    \caption{\mbb{Percentage of estimates that have smaller errors from 2-fold cross-validation from the ANN than the LM. 
    } \label{tab:errors}}
\end{table}

\begin{table}
    \centering
    \begin{tabular}{cllllll}
        & \multicolumn{2}{c}{LM} & \multicolumn{2}{c}{ANN} \\\cline{2-5}\Tstrut
        Variable & CEP & RRL & CEP & RRL \\\hline\hline
        $M/$M$_\odot$ & 0.56 & 0.077 & 0.35 & 0.049 \\
        $\log R/$R$_\odot$ & 0.017 & 0.027 & 0.011 & 0.019 \\
        $\log L/$L$_\odot$ & 0.084 & 0.087 & 0.044 & 0.041 \\
        $T_{\text{eff}}/$K & 209 & 240 & 108 & 98 \\
        $M_I$ & 0.19 & 0.16 & 0.10 & 0.086 \\
        $M_V$ & 0.26 & 0.21 & 0.13 & 0.098 \\
        $V-I$ & 0.070 & 0.056 & 0.033 & 0.028 \\
        $W$ & 0.11 & 0.11 & 0.065 & 0.076 \\\hline
    \end{tabular}
    \caption{\mb{Standard deviations of errors when estimating stellar parameters of Cepheid and RR~Lyrae stars from the LM and the ANN. 
    } \label{tab:significance}}
\end{table}

\subsection{Light Curve Structure Importances}

It is interesting to study which features of the light curve in particular are most useful in constraining fundamental parameters. 
To answer this question we use a separate machine learning technique known as random forests \citep[RFs,][]{breiman2001random}. 
This method works by building an ensemble of decision trees to estimate a function. 
The decision trees consist of `if-then-else' decision rules which are learned from the input data. 
The decision rules are constructed using information theory by considering which input feature is most highly discriminant in constraining the output. 
Thus one way of inferring the relative importance of a given light curve feature is to consider what fraction of decision rules are constructed based on that feature. 
We note that one drawback of this approach is that correlated features share the fraction of the importances, and thus each will appear relatively less important than if only one of the features were part of the analysis. 

\begin{figure*}
    \centering
    \includegraphics[width=0.5\linewidth]{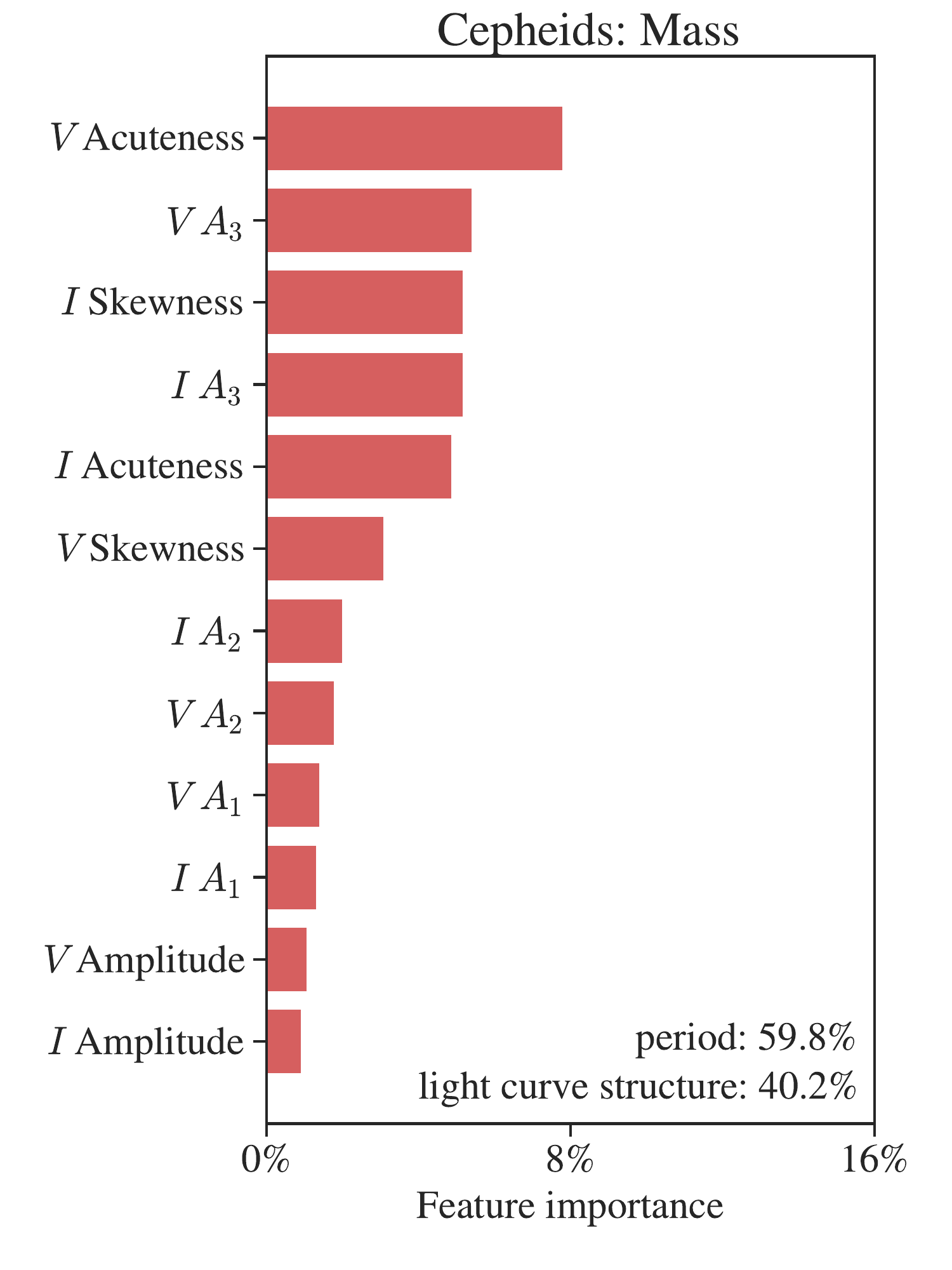}%
    \includegraphics[width=0.5\linewidth]{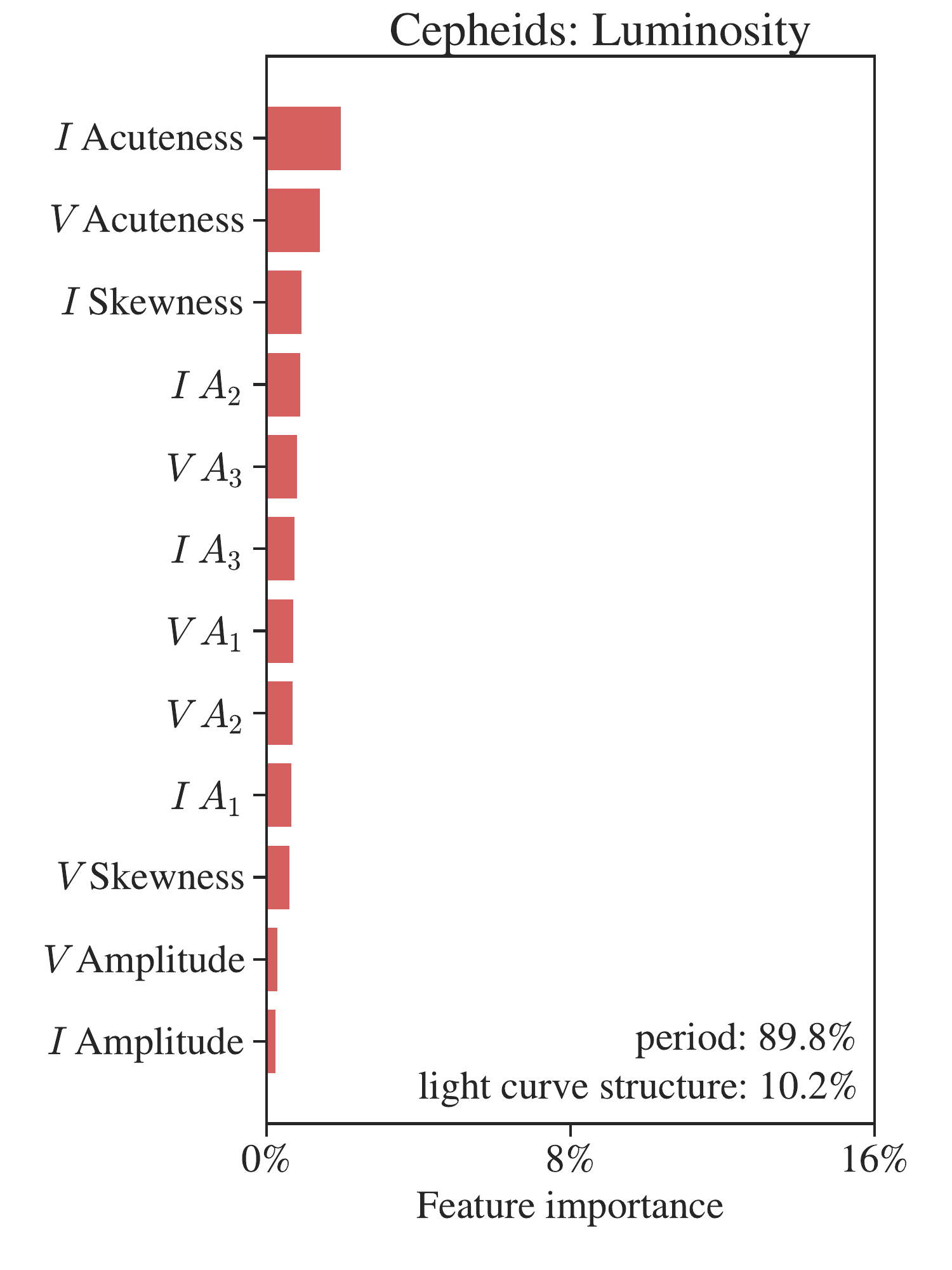}\\%
    \includegraphics[width=0.5\linewidth]{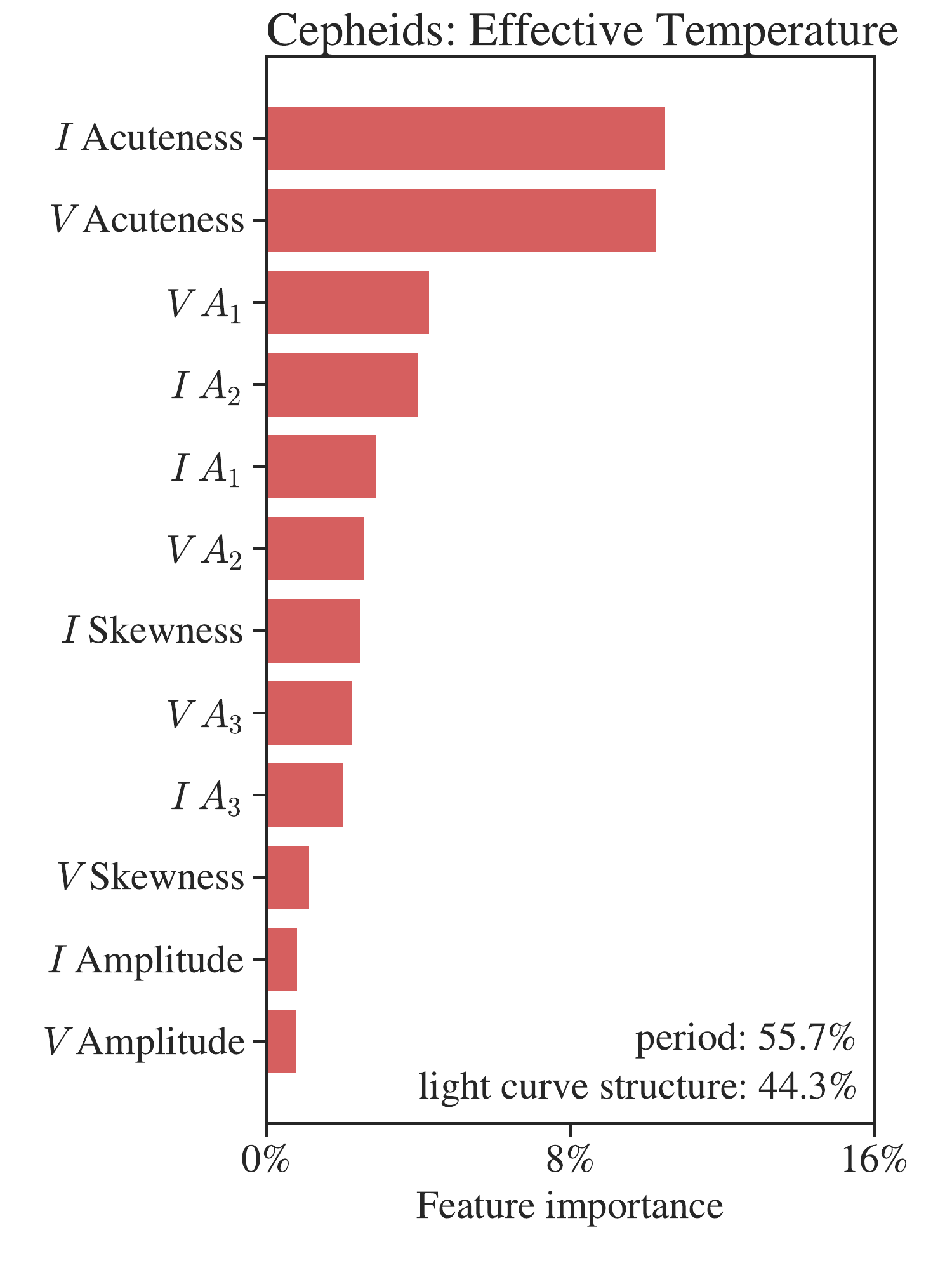}%
    \includegraphics[width=0.5\linewidth]{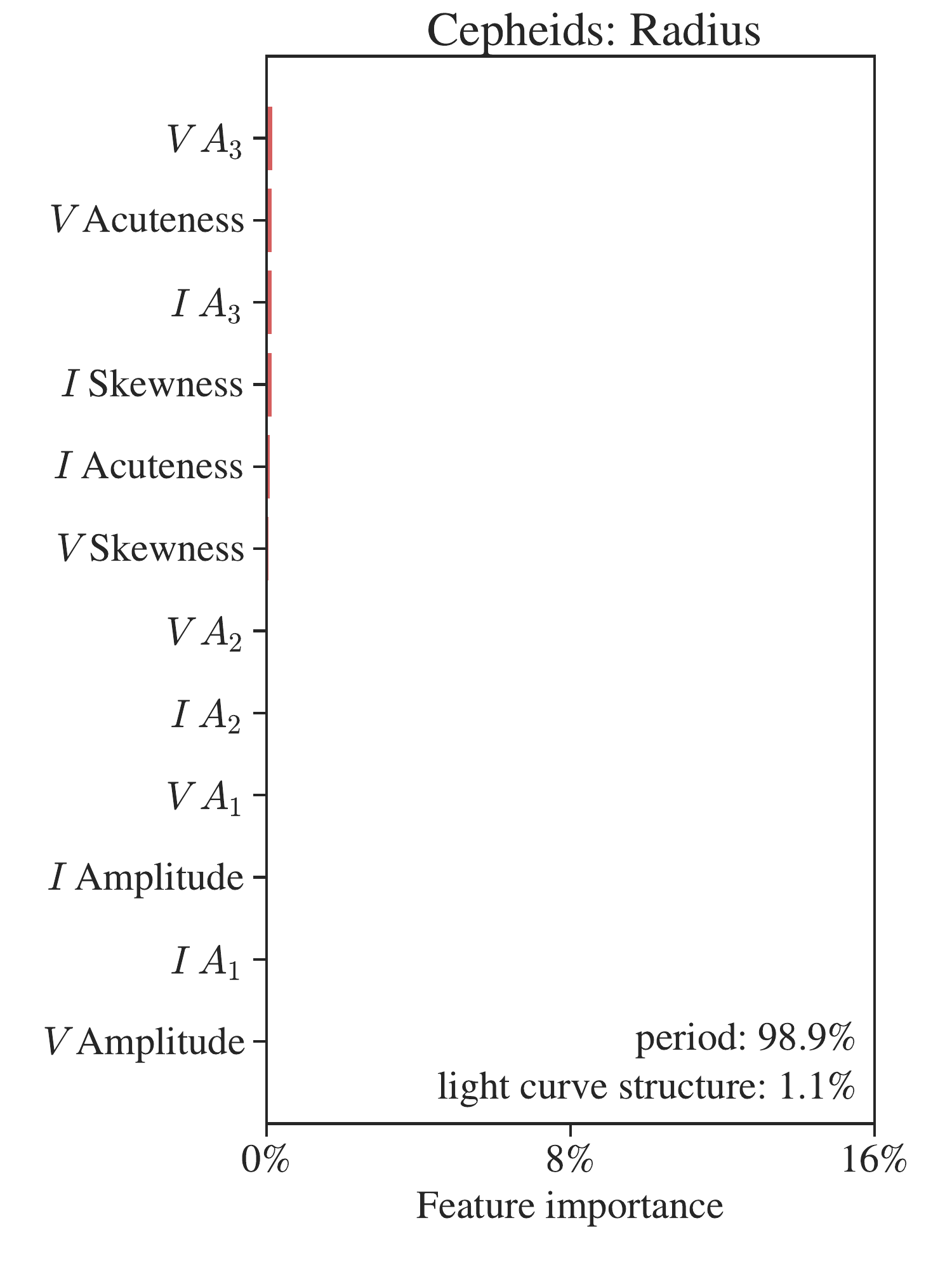}%
    \caption{Feature importances for predicting Cepheid parameters. }
    \label{fig:CEP-importances}
\end{figure*}

\begin{figure*}
    \centering
    \includegraphics[width=0.5\linewidth]{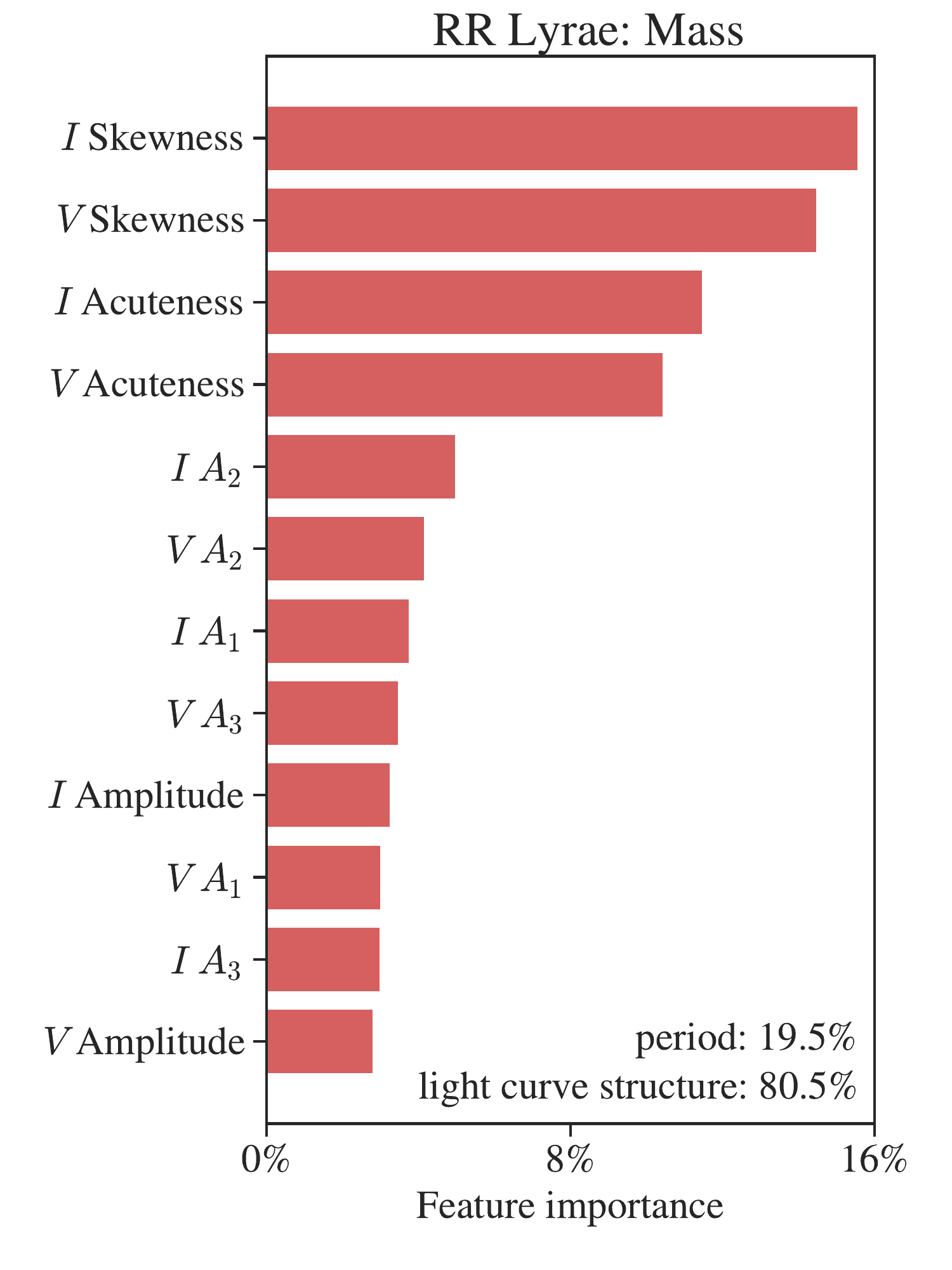}%
    \includegraphics[width=0.5\linewidth]{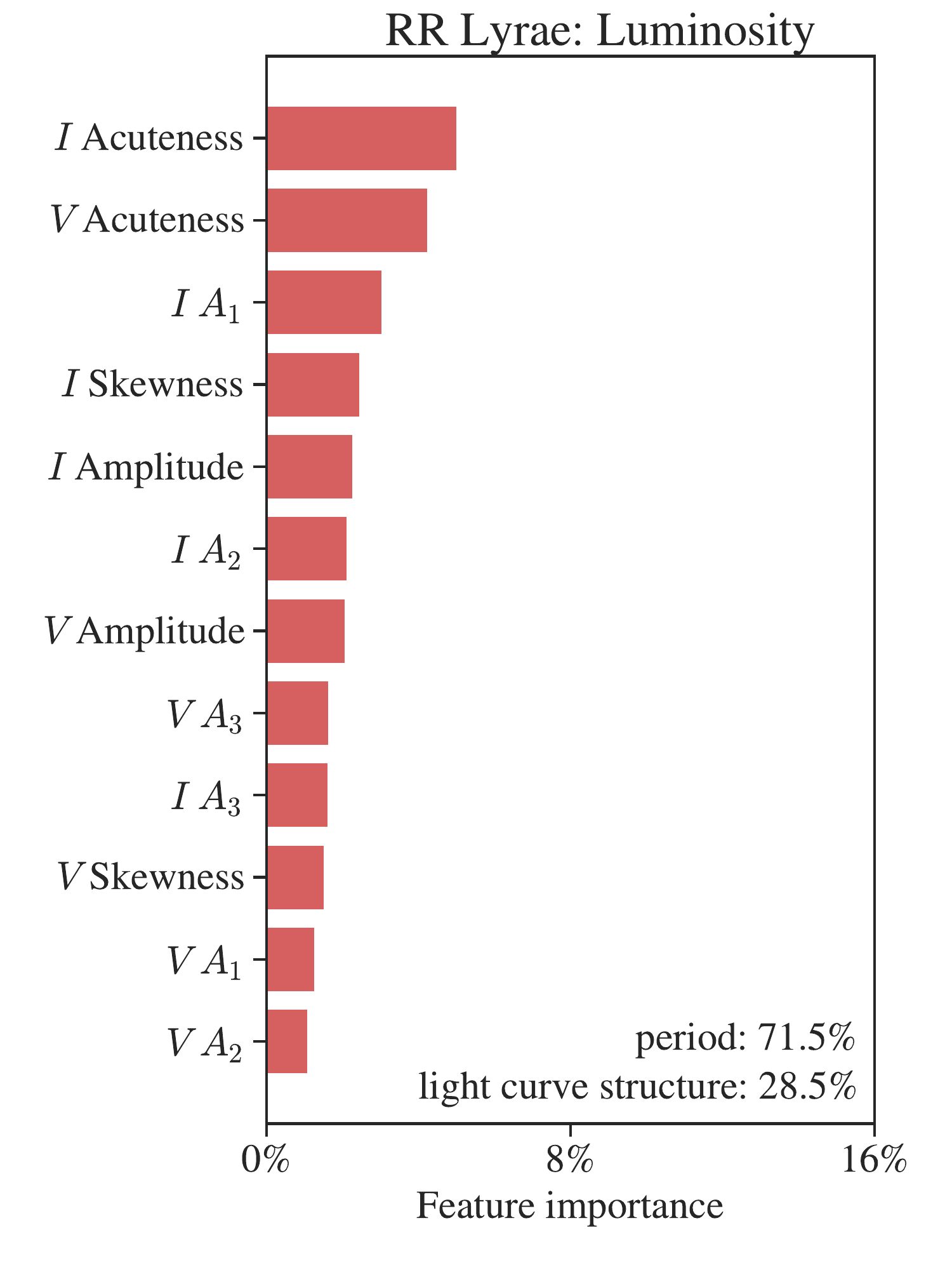}\\%
    \includegraphics[width=0.5\linewidth]{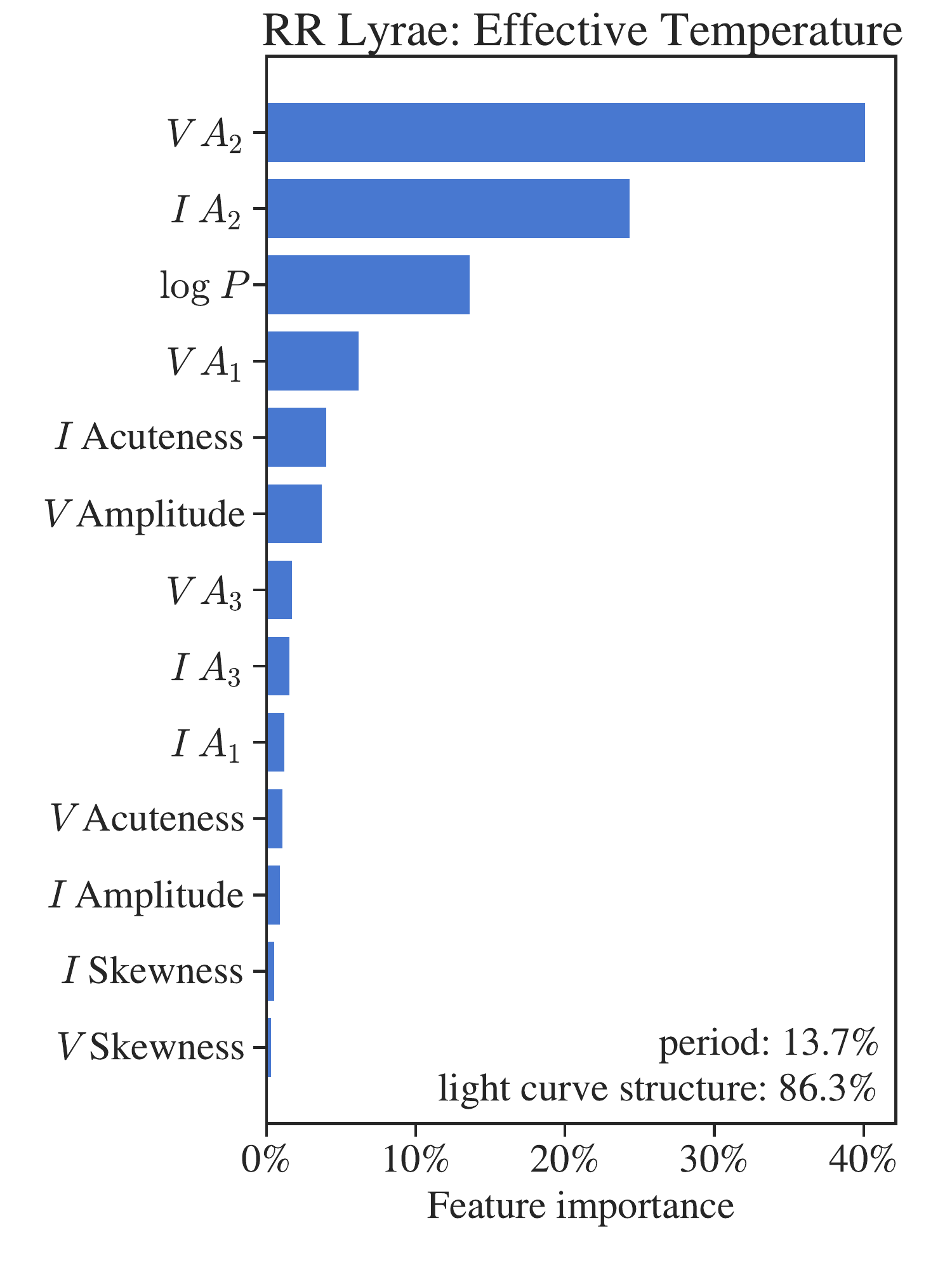}%
    \includegraphics[width=0.5\linewidth]{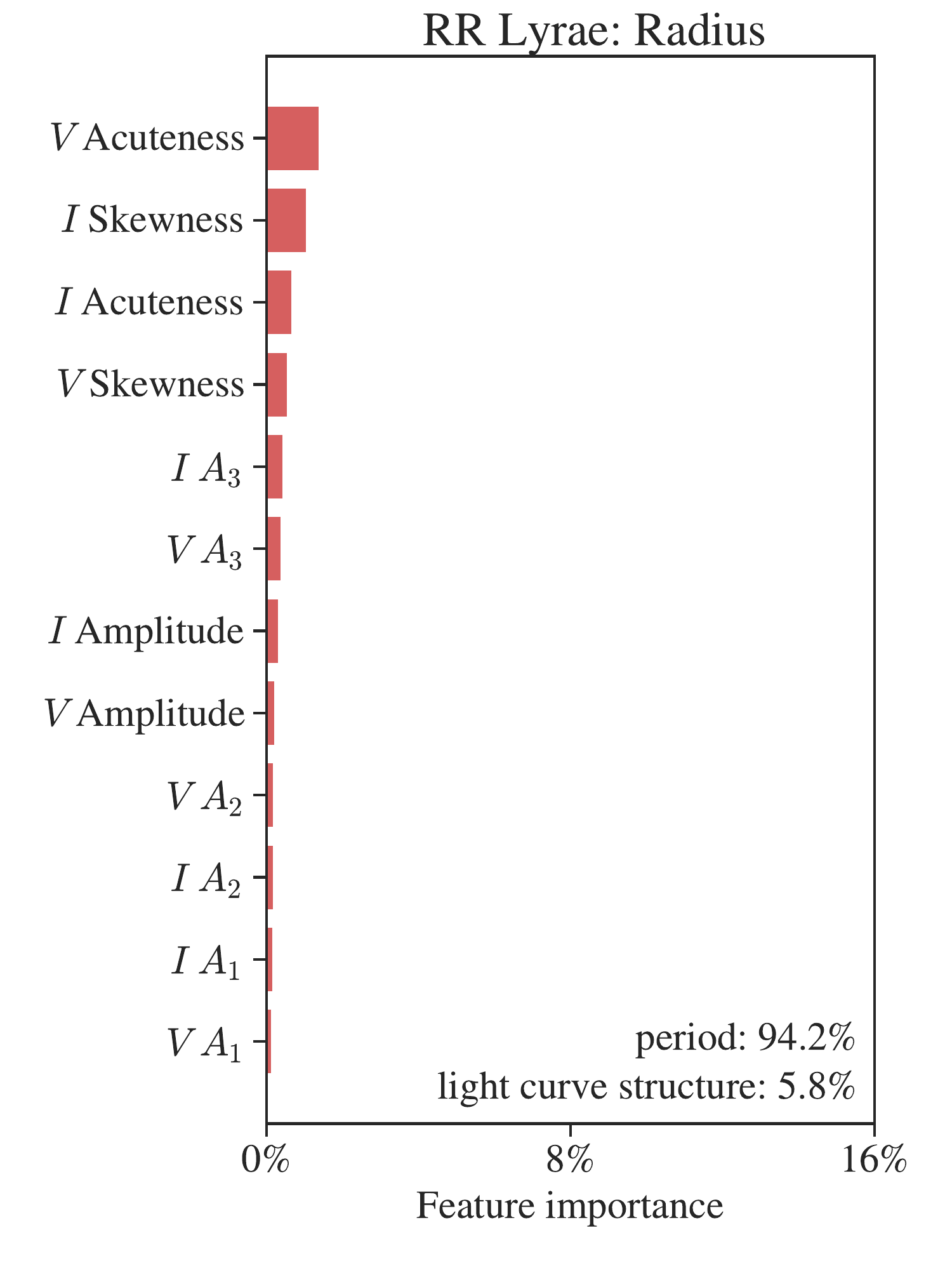}%
    \caption{Feature importances for predicting RR~Lyrae parameters. Note the differences in scale for the $T_{\text{eff}}$ figure. }
    \label{fig:RRL-importances}
\end{figure*}

\begin{figure}
    \centering
    \includegraphics[width=\linewidth]{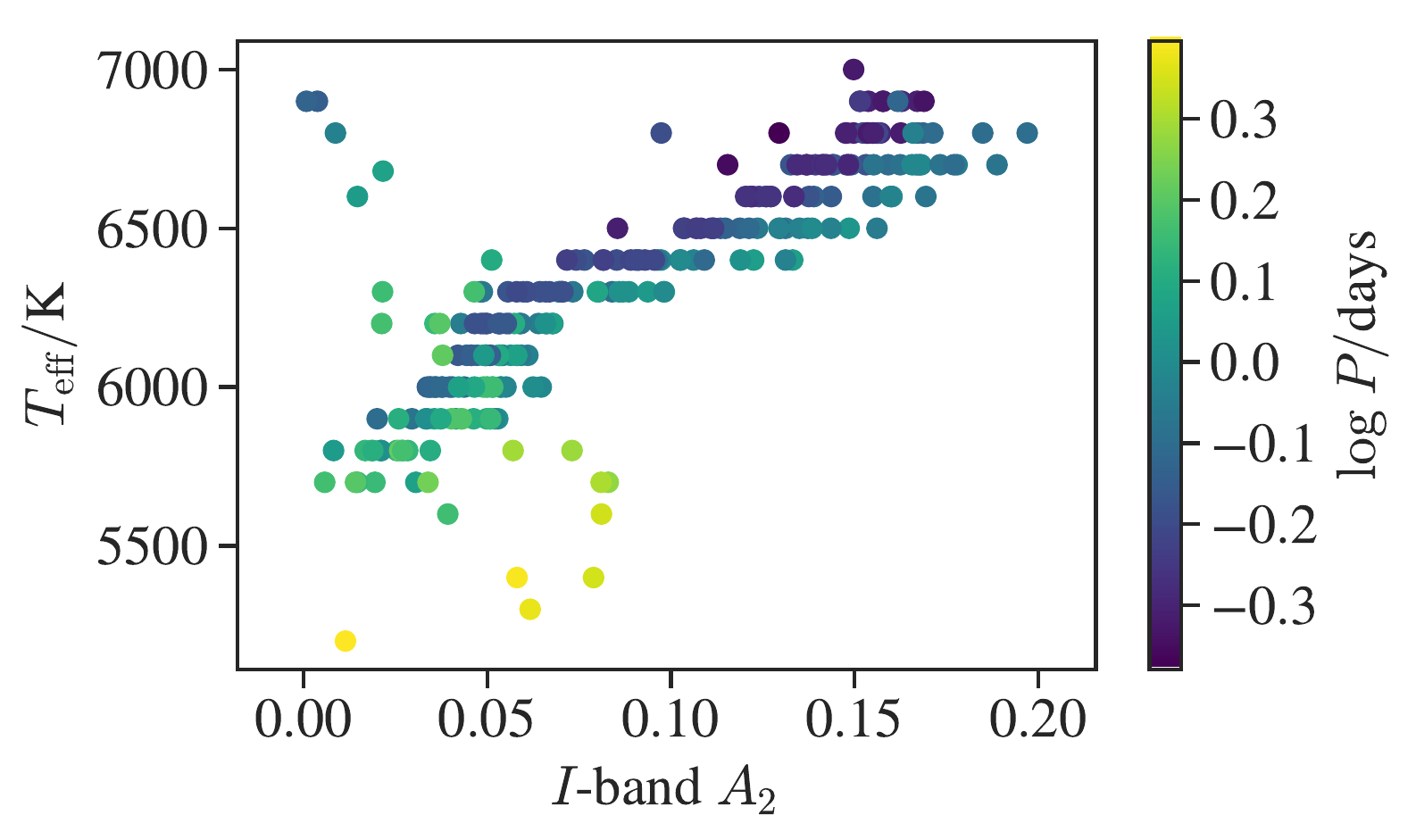}
    \caption{Effective temperatures as a function of the second Fourier amplitude component in the grid of RR~Lyrae models. The same plot for $V$-band amplitudes looks nearly the same.}
    \label{fig:A2Teff}
\end{figure}

Figures~\ref{fig:CEP-importances} and \ref{fig:RRL-importances} display the relative importance of a light curve structure quantity in predicting mass, luminosity, effective temperature and radius for Cepheids and RR~Lyraes respectively. In most cases, the period is the most important followed by either the $V$- or $I$-band acuteness. \citet{1987ApJ...314..252S} had previously noted the importance of light curve acuteness and its possible connection to physics based on one-zone models. However, we note that for RR~Lyrae masses and effective temperatures, the skewnesses and second Fourier amplitude components $A_2$ are either as or more important than the period in predicting that quantity. 
Figure~\ref{fig:A2Teff} confirms this behavior by showing $T_{\text{eff}}$ against $A_2$ amplitudes in the grid of RR~Lyrae models. 
We aim to investigate this more fully in future work. 

In principle we can also use RFs to estimate physical parameters instead of the ANN. 
\mb{However, in this case we found that the ANN had slightly better performance (e.g., an average mean squared error of 0.10 from the ANN as opposed to 0.12 from the RF as measured using 5-fold cross validation on Cepheid masses; contrast with that of 0.29 from an LM).} 
Thus we use the ANN when predicting physical parameters, and use RFs only to illuminate feature importances. 
\mbb{Note that there are also techniques for assessing the importances of features from an ANN, which could in principle give different importances than an RF. 
Nevertheless, this analysis serves to give at least a rough idea of the components of lightcurve structure that are reflective of various stellar parameters. } 

\section{Applications} \label{sec:applications}

In the previous sections, we have used machine learning methods to estimate fundamental stellar parameters using period and light curve structure on theoretical models, and furthermore demonstrated that the addition of light curve structure is statistically important. 
Here we seek to apply these trained ANNs to predict these same quantities, but now using real observed data of Cepheid and RR~Lyrae stars as input instead of models. 
In order to prevent extrapolation, we only analyze stars whose periods and light curve structure parameters are within the inner 90\% of those spanned by the grid of theoretical models, and furthermore exclude any star whose estimated fundamental parameters fall outside the range spanned by the grid of theoretical models. 

\begin{figure*}
    \centering
    \includegraphics[width=0.5\linewidth]{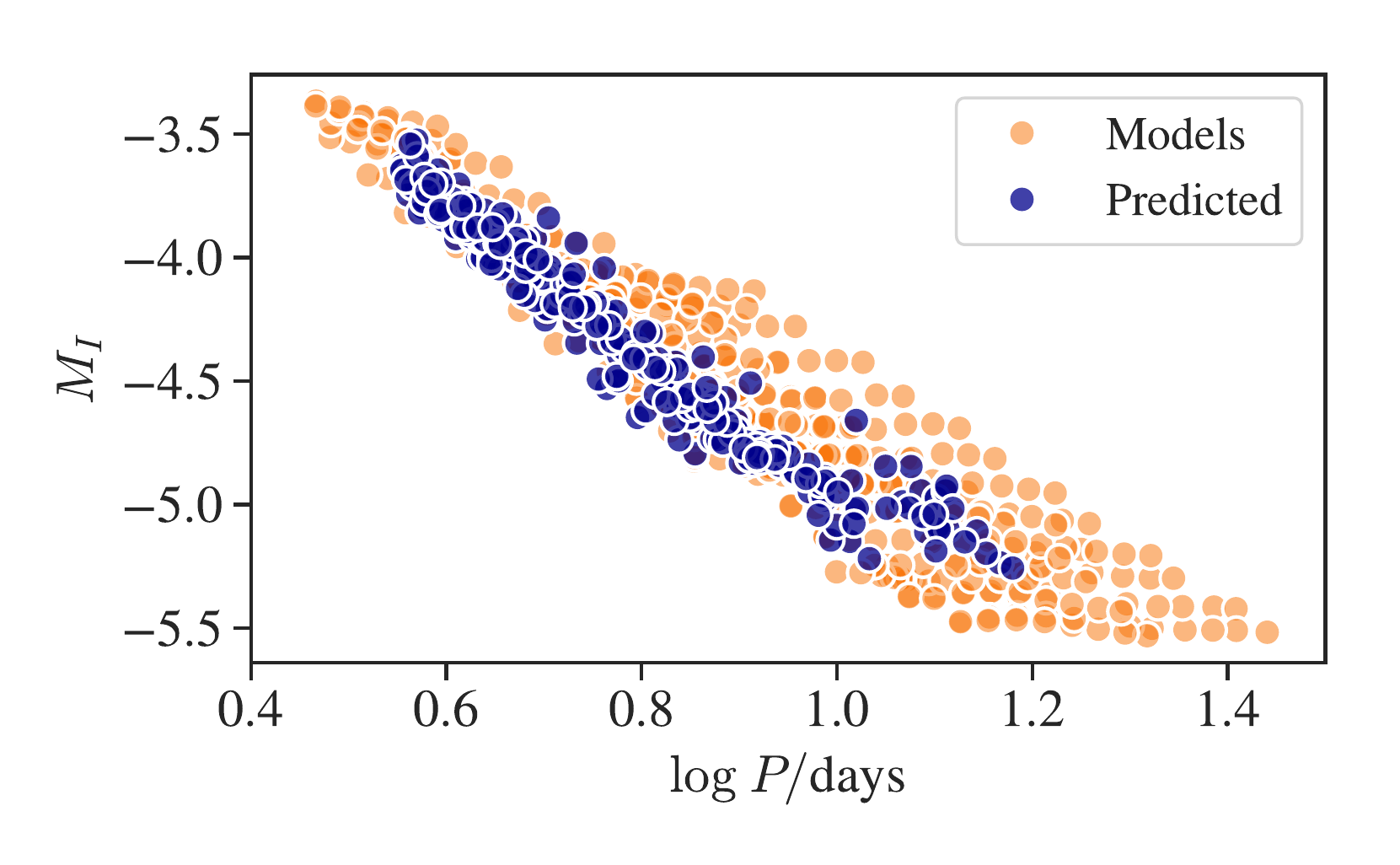}%
    \includegraphics[width=0.5\linewidth]{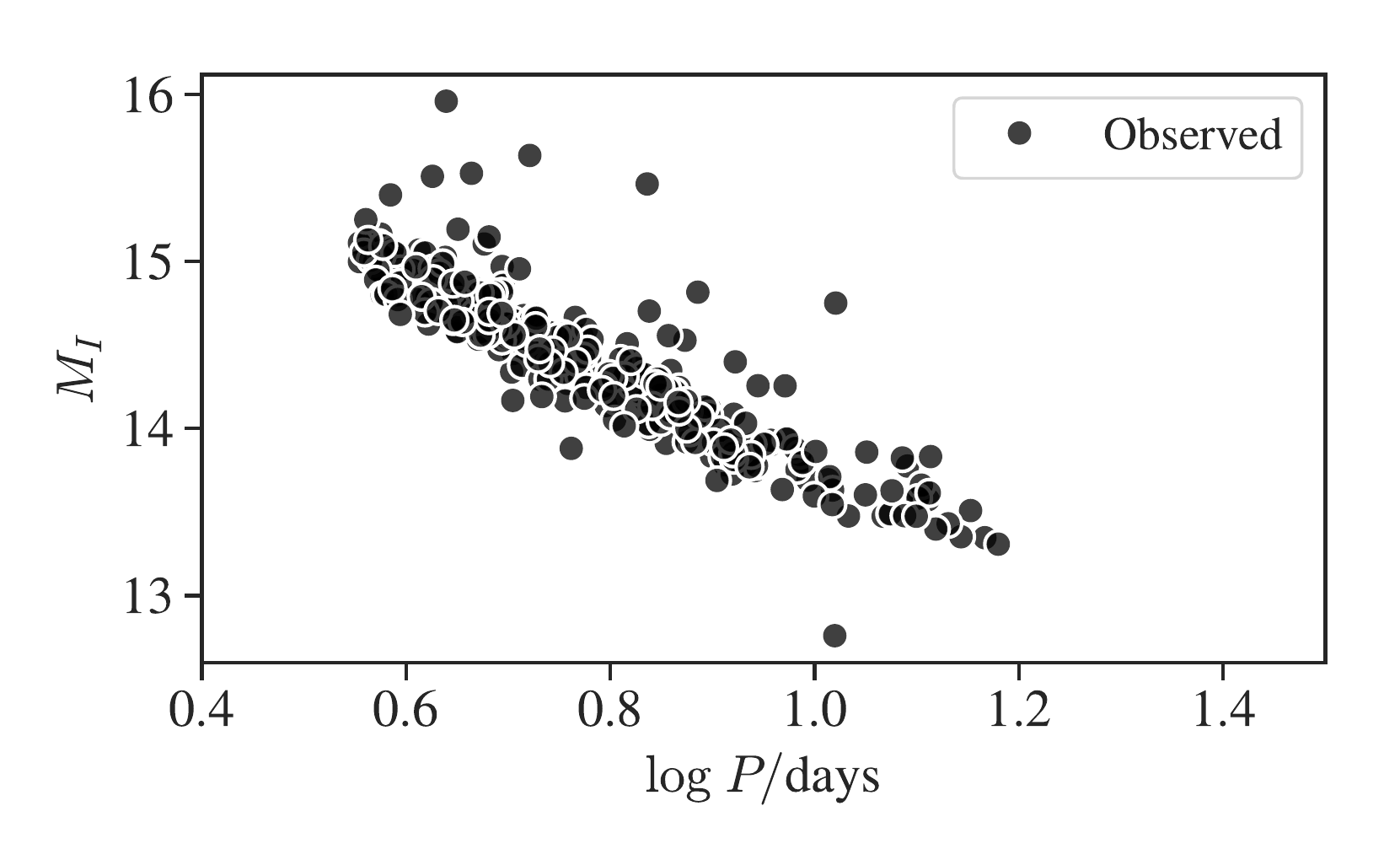}\\%
    \includegraphics[width=0.5\linewidth]{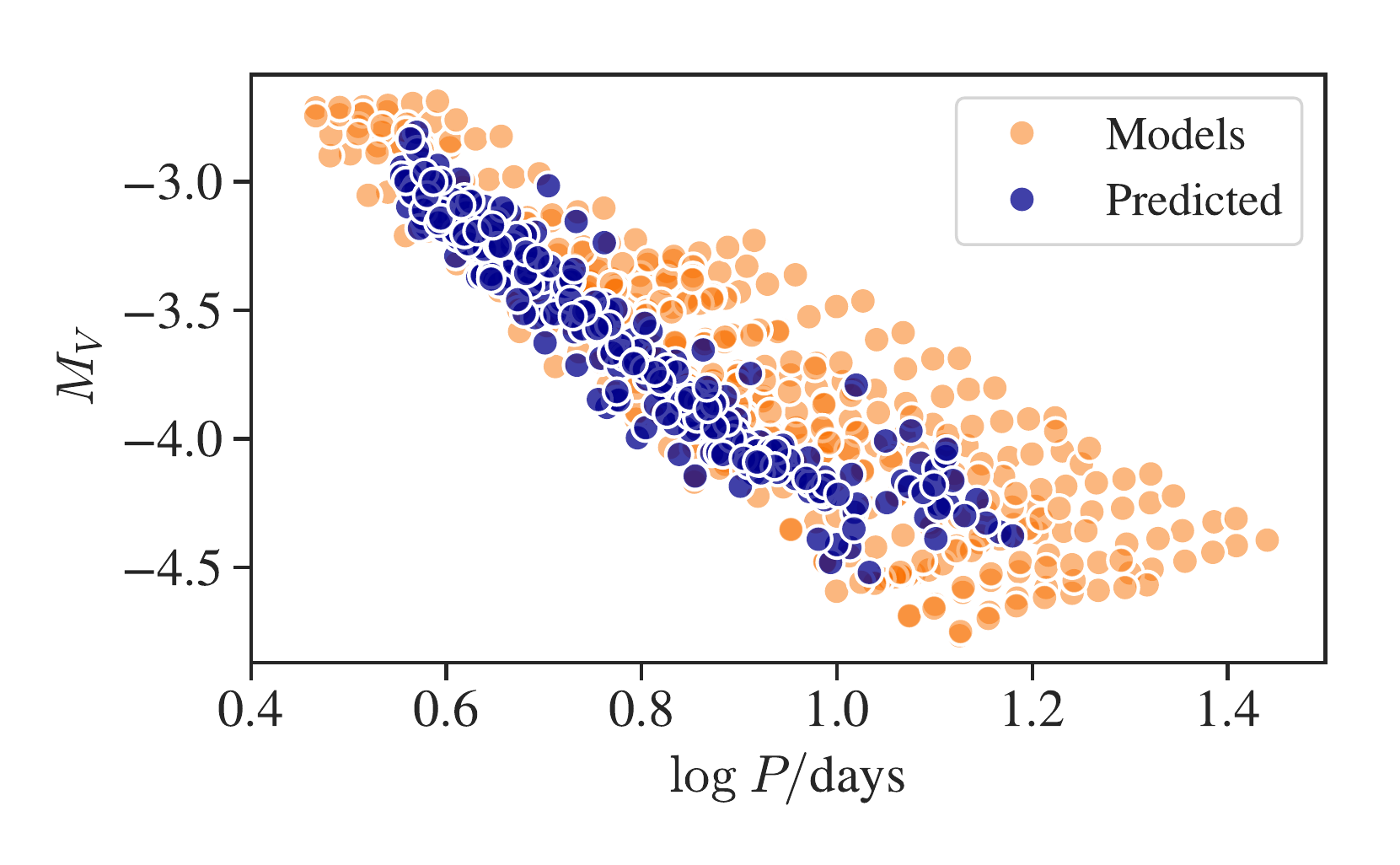}%
    \includegraphics[width=0.5\linewidth]{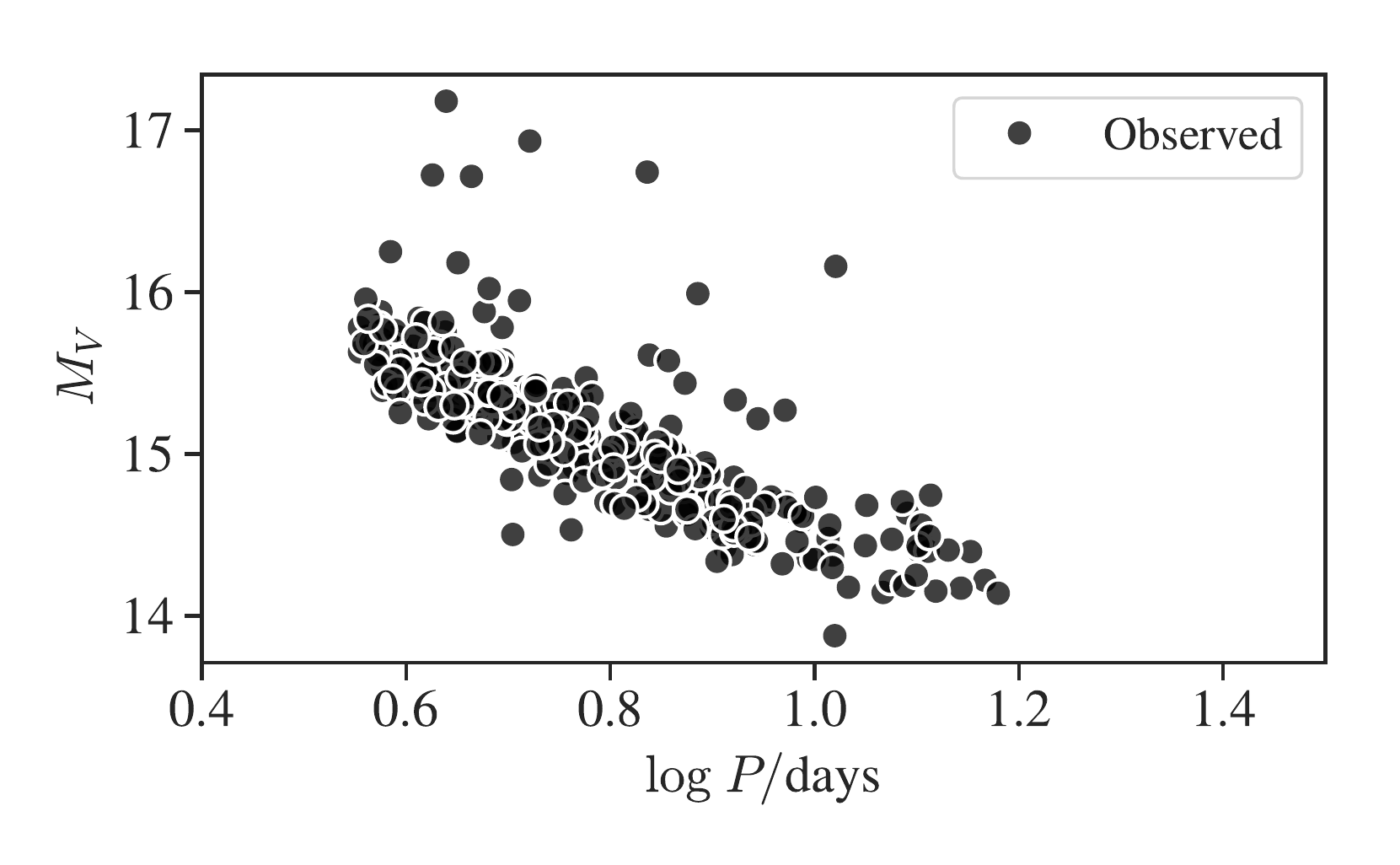}\\%
    \includegraphics[width=0.5\linewidth]{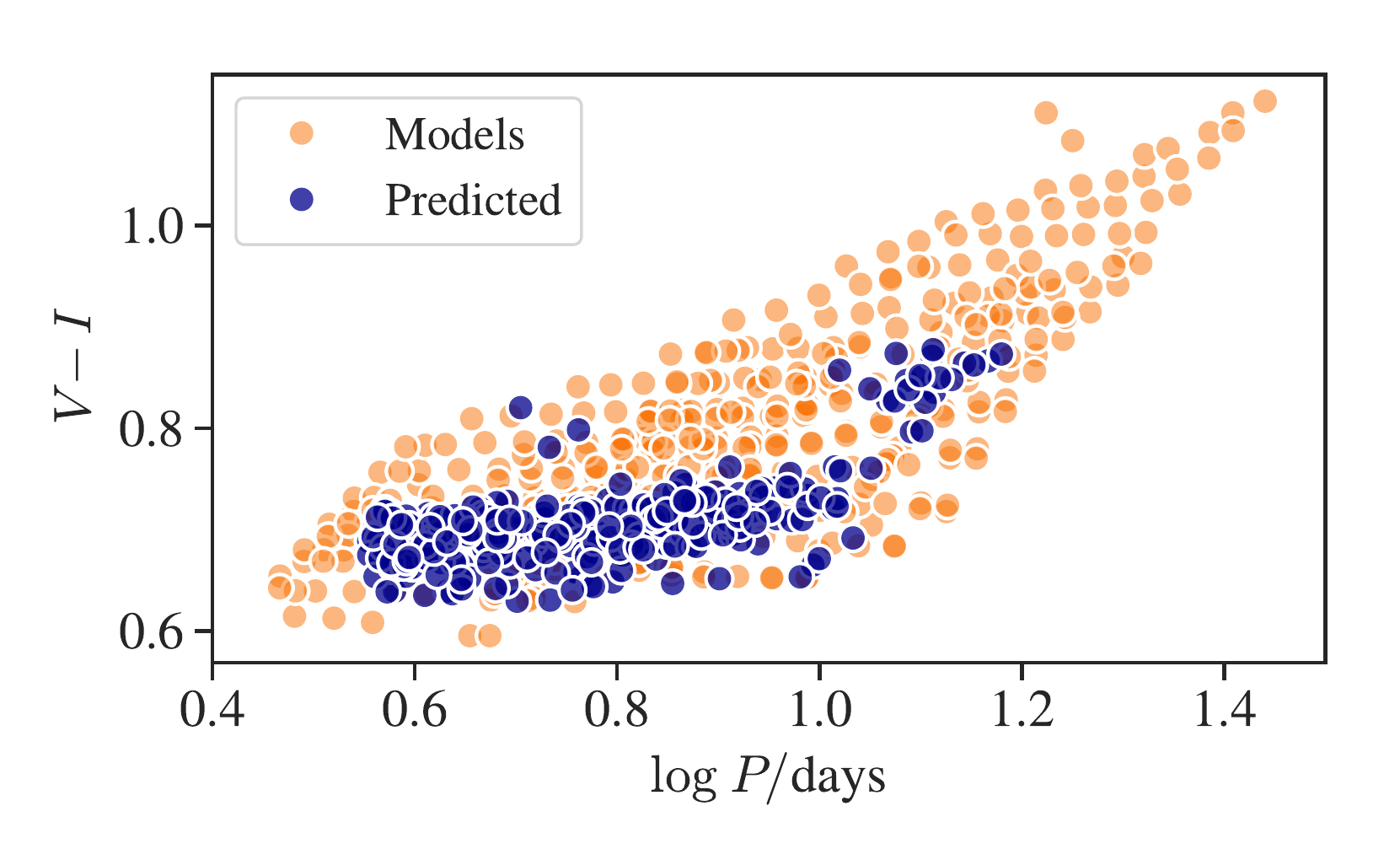}%
    \includegraphics[width=0.5\linewidth]{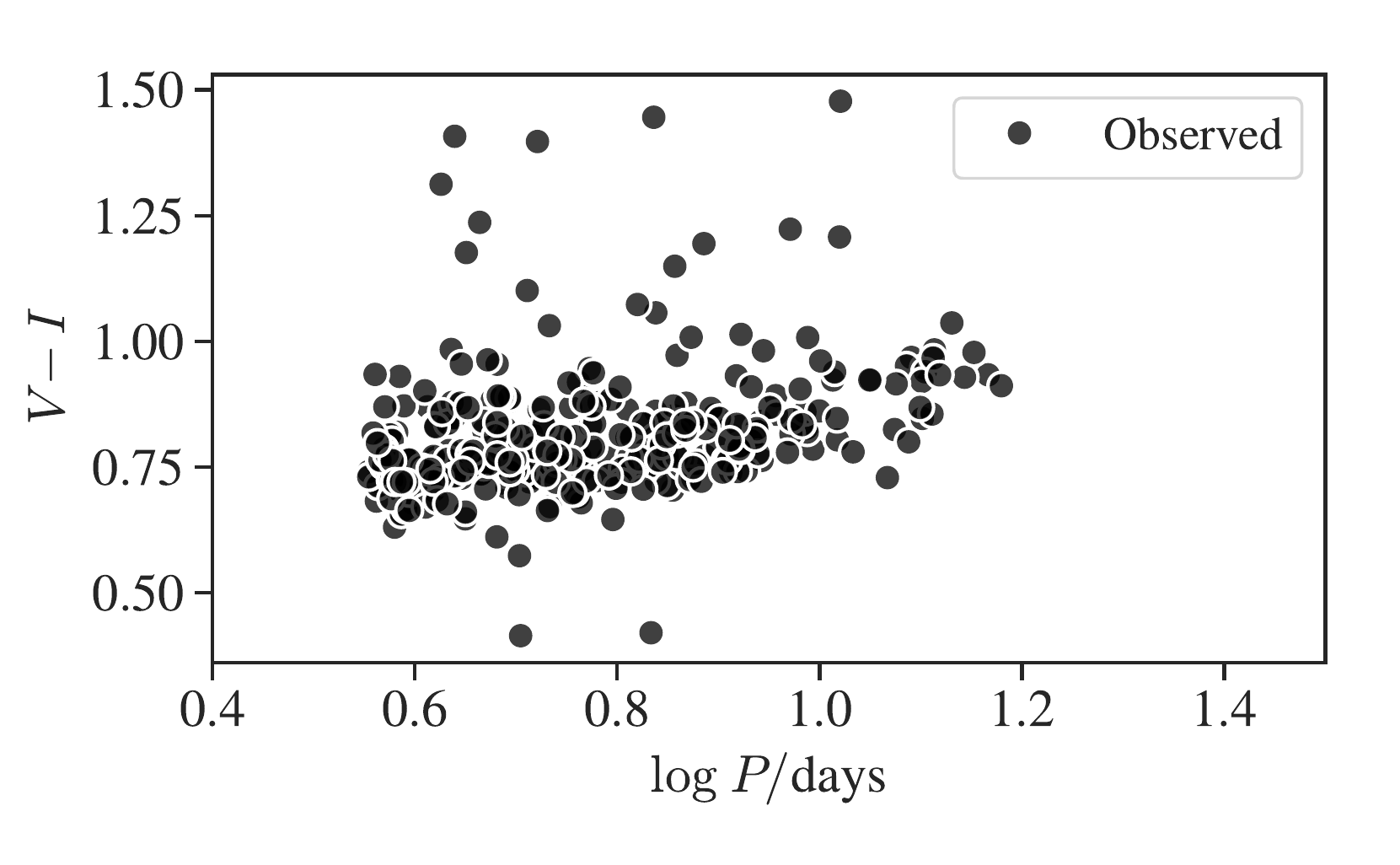}\\%
    \includegraphics[width=0.5\linewidth]{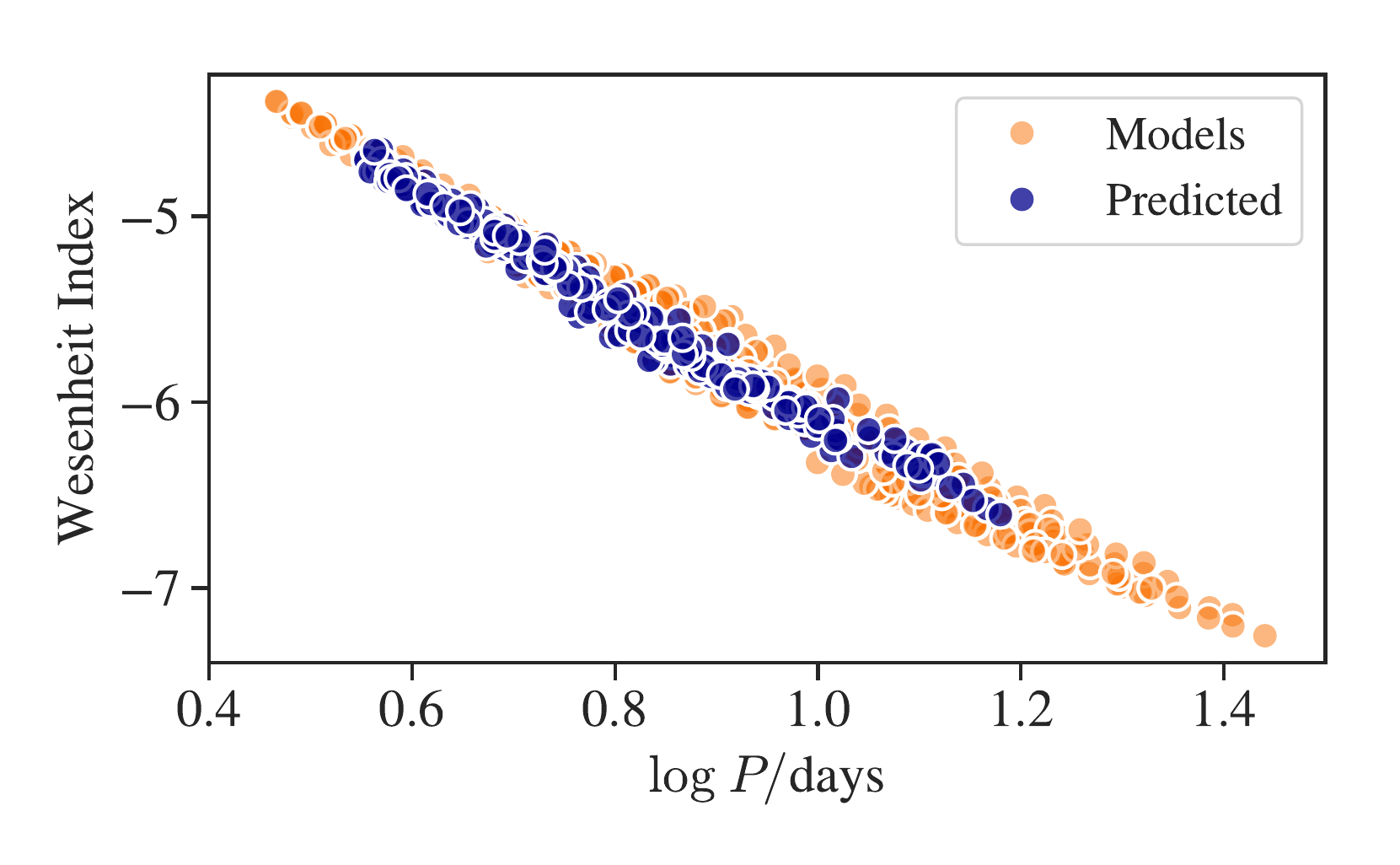}%
    \includegraphics[width=0.5\linewidth]{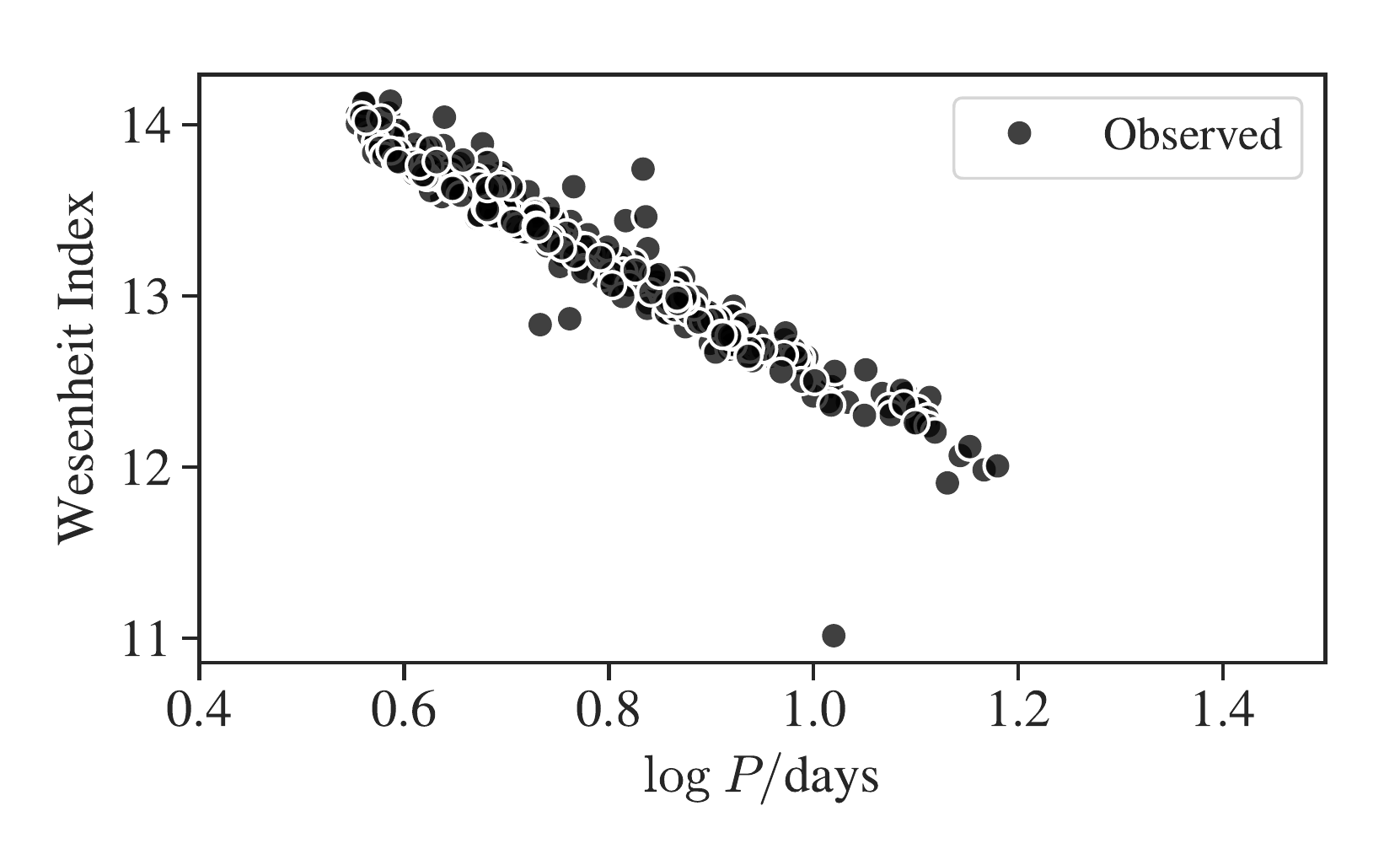}%
    \caption{Theoretical, predicted, and observed period--magnitude, period--color, and period--Wesenheit relations for Cepheids in the LMC. The orange points are the theoretical models. The blue points are estimated magnitudes, colors, and Wesenheit indices made based on period and light curve structure for actual stars. The right panels show the extinction-corrected observed relations for those same stars. } 
    \label{fig:estimates}
\end{figure*}

The results of this analysis are presented in Figure~\ref{fig:estimates}. 
The panels on the left show the theoretical models as well as the results from applying the machine learning model to predict $M_V$, $M_I$, $V-I$, and $W$ for the Cepheid models. 
The panels on the right apply the same machine learning networks to extinction-corrected Cepheid OGLE-IV observations. 
We notice the presence of a kink in the $V$ and $I$ band PL relation at a period of roughly 10 days and a similar feature in the PC relation -- again in both theory and observations \citep[][and references therein]{2016MNRAS.457.1644B}.

\begin{figure*}
    \centering
    \includegraphics[width=0.5\linewidth]{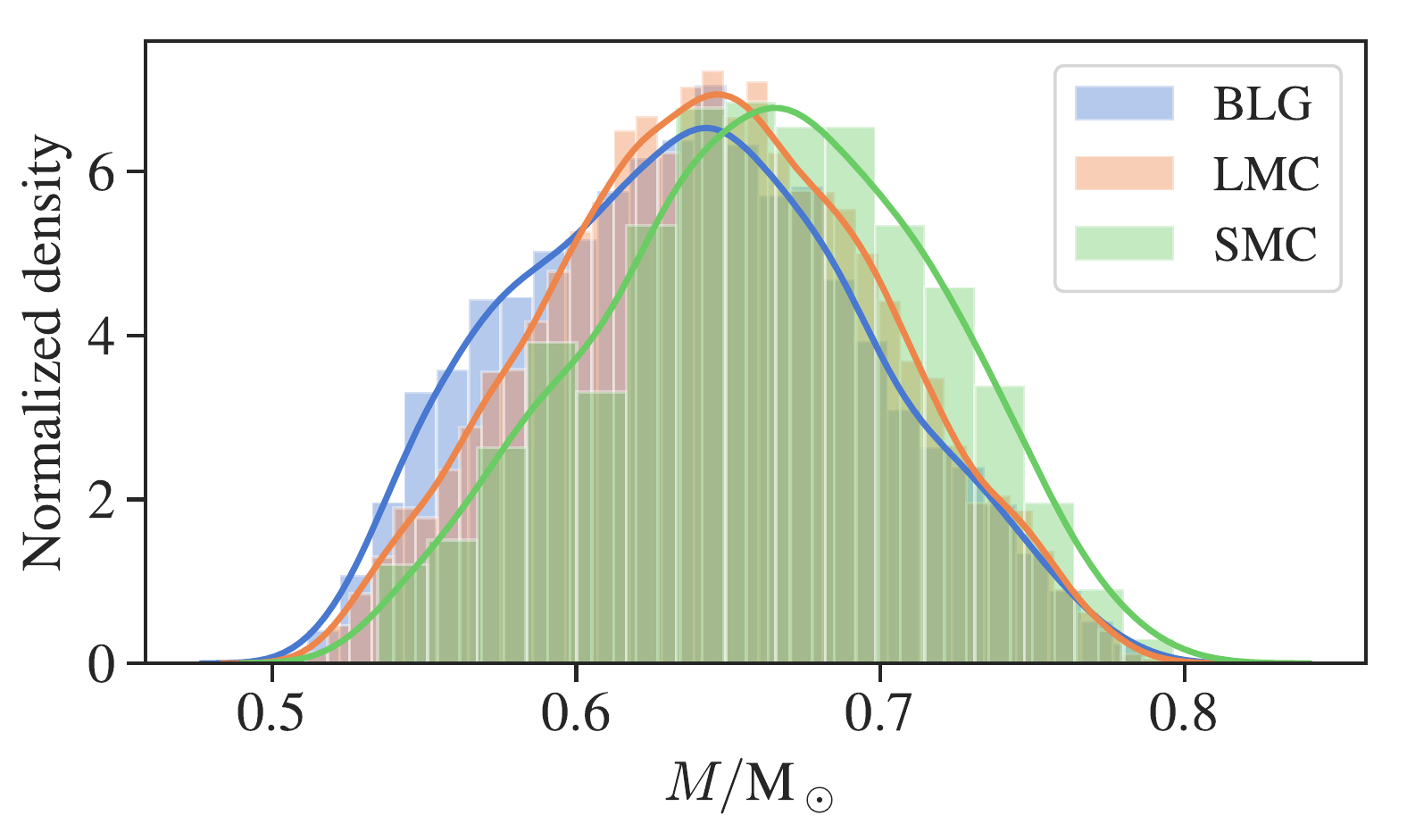}%
    \includegraphics[width=0.5\linewidth]{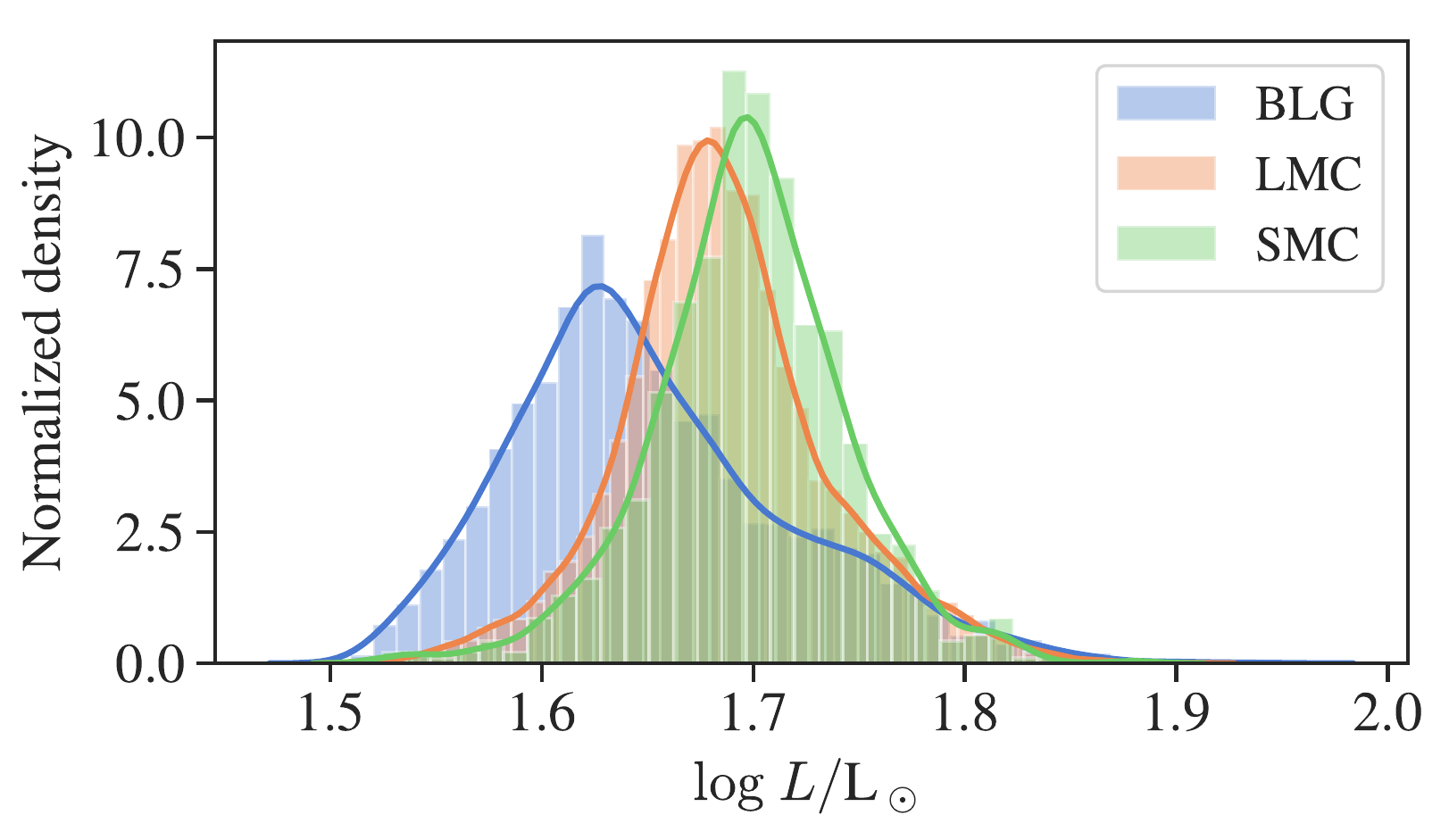}
    \caption{Distributions of estimated masses and luminosities for RR~Lyrae stars in the Galactic bulge (BLG), LMC, and SMC. }
    \label{fig:RRL-distributions}
\end{figure*}

In Figure~\ref{fig:RRL-distributions}, we show estimated RRab masses and luminosities for the Galactic bulge, LMC, and SMC using our machine learning methods. 
Even though the distribution of masses is similar across this range of metallicity, the luminosity distribution, in particular, suggests that metal-poor RRab stars are more luminous. 

\subsection{Distance to the Magellanic Clouds}
\begin{figure}
    \centering  
    \includegraphics[width=\linewidth,trim={0 0.6in 0 0}, clip]{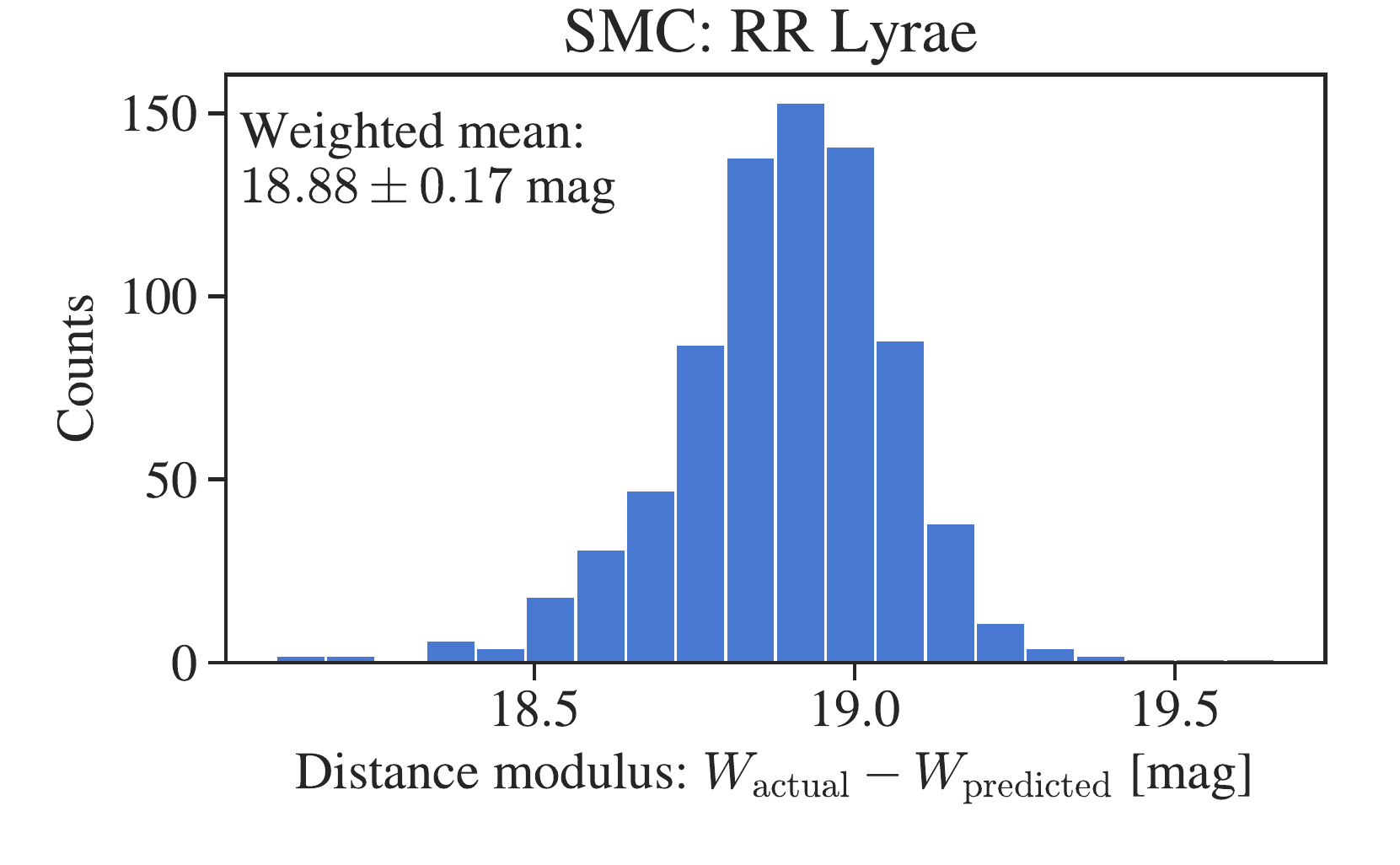}\\[1.5\baselineskip]%
    \includegraphics[width=\linewidth,trim={0 0.6in 0 0}, clip]{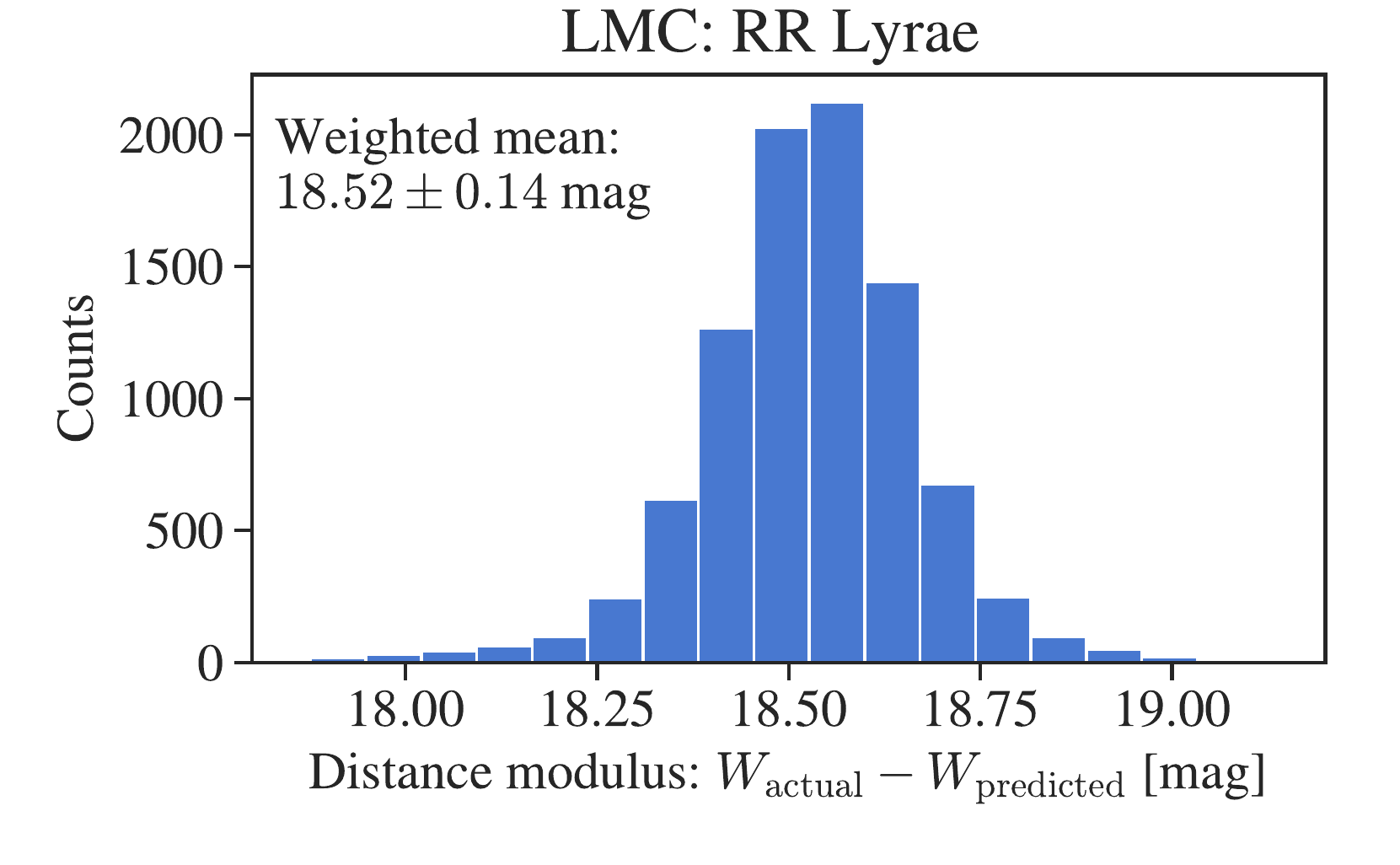}\\[1.5\baselineskip]%
    \includegraphics[width=\linewidth]{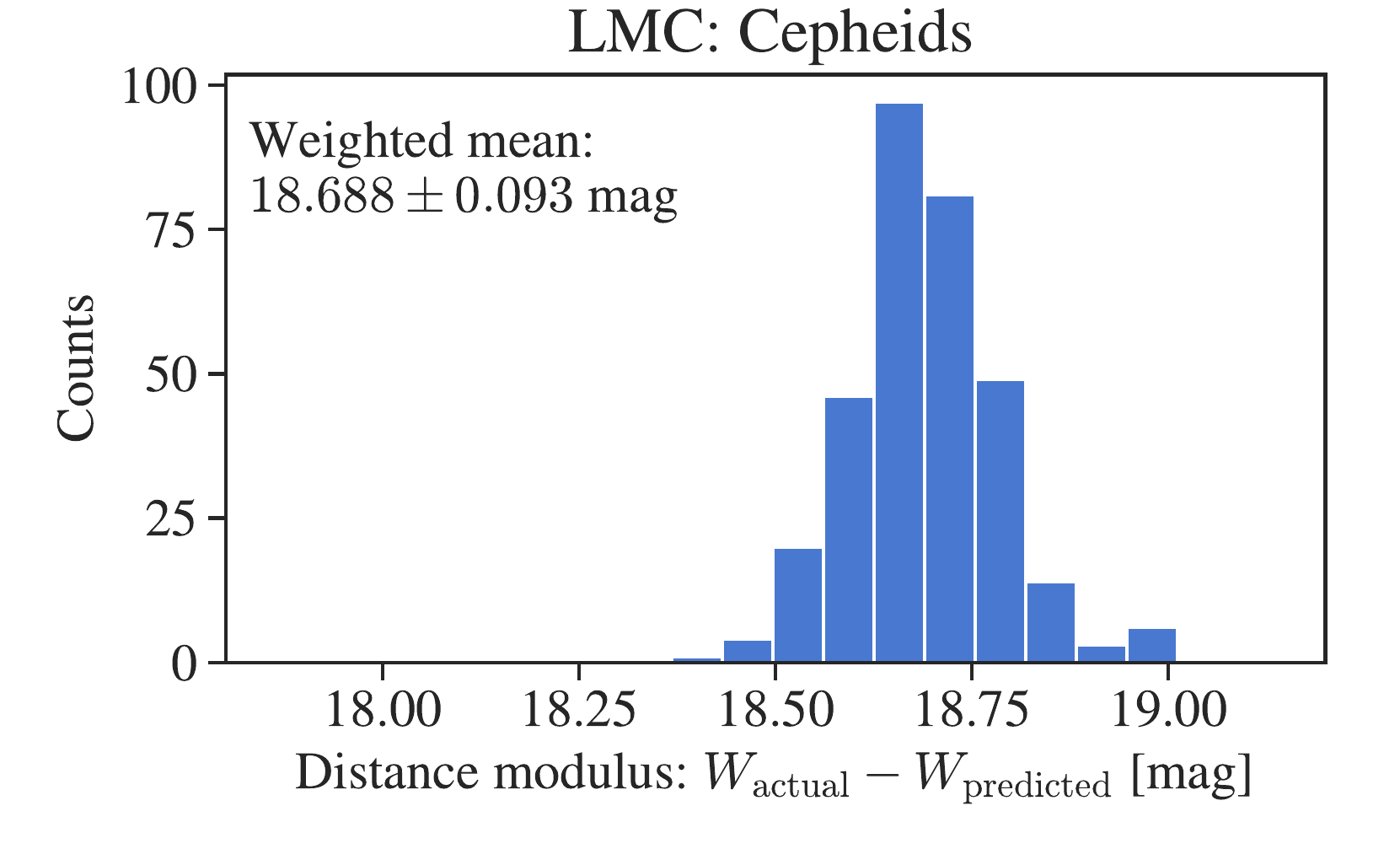}%
    \caption{Histogram of differences between the actual and estimated Wesenheit indices for RR ~Lyrae stars in the SMC (top panel) and RR~Lyrae and Cepheid and stars in the LMC (middle and bottom panels). \label{fig:distance-modulus}}
\end{figure}

We can now apply this method to compute Wesenheit-based distance moduli $W_m - W_M$, where $W_m$ is the observed Wesenheit index and $W_M$ is the estimated one, to each individual star. 
These results are given in Figure~\ref{fig:distance-modulus}. 
After sigma-clipping outliers at the 5$\sigma$ level, a weighted average of these distance moduli yields an LMC distance of ${18.688 \pm 0.093}$~mag from Cepheids and ${18.52 \pm 0.14}$~mag from RR~Lyrae stars. 
We note that these estimates are based on an ANN trained on both the canonical and the non-canonical models. 
Similarly, we obtain an SMC distance of ${18.88 \pm 0.17}$~mag from RR~Lyrae stars. 
These estimates are statistically consistent with those from eclipsing binaries \citep[$\mu_{\text{LMC}} = 18.476 \pm 0.024$~mag, $\mu_{\text{SMC}} = 18.95 \pm 0.07$~mag,][]{2014ApJ...780...59G, 2019Natur.567..200P}. 
With this approach we do not assume any kind of period--Wesenheit relation to make these estimates as is commonly done, though that approach gives approximately the same result, albeit with somewhat higher uncertainty. 

\subsection{Catalogs}
Tables~\ref{tab:LMC-CEP}, \ref{tab:LMC-RRL}, \ref{tab:SMC-RRL}, and~\ref{tab:BLG-RRL} lists the estimated fundamental parameters for Cepheids in the LMC and RR Lyraes in the LMC, SMC, and Galactic Bulge, respectively. 
The uncertainties on these estimates are obtained by perturbing the observed light curve parameters 100 times and passing each random realization of noise through the ANN. 
The resulting random uncertainties are then added in quadrature with the systematic uncertainties (\emph{cf.}~Table~\ref{tab:significance}) to obtain the stated total uncertainty. 
\mb{Note that as here we have perturbed the light curve parameters themselves and not the underlying light curves, we neglect correlations in the the light curve parameters. } 

\begin{table*}\hspace*{-0.6cm}
    \centering
\begin{tabular}{ccccccccc}\hline
                ID &         $M/$M$_\odot$ &      $\log R/$R$_\odot$ &    $\log L/$L$_\odot$ &    $T_{\text{eff}}/K$  & $M_I$ &    $M_V$ & $V-I$ &  $W$ \\\hline\hline
0033 & 5.98 $\pm$ 0.39 & 1.756 $\pm$ 0.014 & 3.513 $\pm$ 0.057 & 5780 $\pm$ 140 & -4.66 $\pm$ 0.13 & -3.96 $\pm$ 0.17 & 0.700 $\pm$ 0.044 & -5.748 $\pm$ 0.083 \\
0050 & 6.16 $\pm$ 0.40 & 1.780 $\pm$ 0.014 & 3.541 $\pm$ 0.059 & 5710 $\pm$ 140 & -4.74 $\pm$ 0.13 & -4.02 $\pm$ 0.18 & 0.720 $\pm$ 0.046 & -5.862 $\pm$ 0.082 \\
0057 & 6.87 $\pm$ 0.40 & 1.861 $\pm$ 0.012 & 3.693 $\pm$ 0.045 & 5680 $\pm$ 110 & -5.14 $\pm$ 0.10 & -4.42 $\pm$ 0.13 & 0.721 $\pm$ 0.035 & -6.260 $\pm$ 0.066 \\
0078 & 5.69 $\pm$ 0.37 & 1.749 $\pm$ 0.012 & 3.493 $\pm$ 0.045 & 5760 $\pm$ 110 & -4.60 $\pm$ 0.10 & -3.90 $\pm$ 0.13 & 0.708 $\pm$ 0.034 & -5.707 $\pm$ 0.068 \\
0103 & 6.33 $\pm$ 0.40 & 1.760 $\pm$ 0.014 & 3.570 $\pm$ 0.049 & 5940 $\pm$ 110 & -4.79 $\pm$ 0.11 & -4.14 $\pm$ 0.14 & 0.647 $\pm$ 0.034 & -5.795 $\pm$ 0.077 \\
0107 & 5.98 $\pm$ 0.36 & 1.799 $\pm$ 0.012 & 3.584 $\pm$ 0.051 & 5730 $\pm$ 120 & -4.84 $\pm$ 0.11 & -4.12 $\pm$ 0.15 & 0.720 $\pm$ 0.039 & -5.961 $\pm$ 0.070 \\
0123 & 5.82 $\pm$ 0.36 & 1.736 $\pm$ 0.012 & 3.461 $\pm$ 0.046 & 5730 $\pm$ 110 & -4.52 $\pm$ 0.11 & -3.81 $\pm$ 0.14 & 0.716 $\pm$ 0.035 & -5.639 $\pm$ 0.069 \\
0155 & 6.25 $\pm$ 0.39 & 1.835 $\pm$ 0.012 & 3.702 $\pm$ 0.045 & 5870 $\pm$ 110 & -5.14 $\pm$ 0.10 & -4.48 $\pm$ 0.13 & 0.665 $\pm$ 0.034 & -6.182 $\pm$ 0.069 \\
0158 & 5.48 $\pm$ 0.38 & 1.735 $\pm$ 0.013 & 3.476 $\pm$ 0.048 & 5800 $\pm$ 110 & -4.55 $\pm$ 0.11 & -3.86 $\pm$ 0.14 & 0.698 $\pm$ 0.035 & -5.643 $\pm$ 0.073 \\
0162 & 5.46 $\pm$ 0.51 & 1.647 $\pm$ 0.019 & 3.367 $\pm$ 0.053 & 6010 $\pm$ 110 & -4.25 $\pm$ 0.13 & -3.62 $\pm$ 0.15 & 0.629 $\pm$ 0.034 & -5.23 $\pm$ 0.10 \\\hline
\end{tabular}
    \caption{Masses, radii, luminosities, effective temperatures, $I$- and $V$-band magnitudes, color, and Wesenheit indices estimated using machine learning for Cepheids in the LMC. Only 10 rows are shown; the remaining are available online in a machine-readable format.  \label{tab:LMC-CEP}}
\end{table*}

\begin{table*}\hspace*{-0.6cm}
    \centering
\begin{tabular}{ccccccccc}\hline
                ID &         $M/$M$_\odot$ &      $\log R/$R$_\odot$ &    $\log L/$L$_\odot$ &    $T_{\text{eff}}/K$  & $M_I$ &    $M_V$ & $V-I$ &  $W$ \\\hline\hline
00008 & 0.605 $\pm$ 0.054 & 0.787 $\pm$ 0.022 & 1.778 $\pm$ 0.044 & 6490 $\pm$ 110 & -0.164 $\pm$ 0.091 & 0.33 $\pm$ 0.10 & 0.508 $\pm$ 0.031 & -0.964 $\pm$ 0.086 \\
00010 & 0.676 $\pm$ 0.059 & 0.746 $\pm$ 0.022 & 1.667 $\pm$ 0.044 & 6380 $\pm$ 110 & 0.140 $\pm$ 0.091 & 0.67 $\pm$ 0.10 & 0.542 $\pm$ 0.030 & -0.714 $\pm$ 0.087 \\
00027 & 0.714 $\pm$ 0.056 & 0.843 $\pm$ 0.022 & 1.828 $\pm$ 0.045 & 6270 $\pm$ 110 & -0.284 $\pm$ 0.093 & 0.27 $\pm$ 0.10 & 0.567 $\pm$ 0.031 & -1.166 $\pm$ 0.089 \\
00040 & 0.676 $\pm$ 0.056 & 0.757 $\pm$ 0.021 & 1.665 $\pm$ 0.045 & 6310 $\pm$ 110 & 0.135 $\pm$ 0.092 & 0.68 $\pm$ 0.10 & 0.560 $\pm$ 0.030 & -0.748 $\pm$ 0.084 \\
00072 & 0.693 $\pm$ 0.055 & 0.767 $\pm$ 0.022 & 1.698 $\pm$ 0.045 & 6360 $\pm$ 110 & 0.061 $\pm$ 0.093 & 0.59 $\pm$ 0.11 & 0.546 $\pm$ 0.030 & -0.791 $\pm$ 0.084 \\
00079 & 0.623 $\pm$ 0.055 & 0.715 $\pm$ 0.021 & 1.660 $\pm$ 0.044 & 6590 $\pm$ 110 & 0.185 $\pm$ 0.091 & 0.68 $\pm$ 0.10 & 0.498 $\pm$ 0.029 & -0.591 $\pm$ 0.083 \\
00082 & 0.665 $\pm$ 0.058 & 0.729 $\pm$ 0.021 & 1.715 $\pm$ 0.043 & 6690 $\pm$ 110 & 0.071 $\pm$ 0.089 & 0.55 $\pm$ 0.10 & 0.478 $\pm$ 0.029 & -0.669 $\pm$ 0.083 \\
00103 & 0.653 $\pm$ 0.061 & 0.752 $\pm$ 0.023 & 1.696 $\pm$ 0.044 & 6450 $\pm$ 110 & 0.076 $\pm$ 0.091 & 0.60 $\pm$ 0.10 & 0.529 $\pm$ 0.030 & -0.744 $\pm$ 0.089 \\
00112 & 0.637 $\pm$ 0.066 & 0.728 $\pm$ 0.024 & 1.692 $\pm$ 0.045 & 6610 $\pm$ 120 & 0.108 $\pm$ 0.093 & 0.60 $\pm$ 0.10 & 0.495 $\pm$ 0.032 & -0.662 $\pm$ 0.094 \\
00118 & 0.649 $\pm$ 0.056 & 0.736 $\pm$ 0.020 & 1.690 $\pm$ 0.043 & 6550 $\pm$ 100 & 0.093 $\pm$ 0.088 & 0.617 $\pm$ 0.099 & 0.511 $\pm$ 0.029 & -0.697 $\pm$ 0.081 \\\hline
\end{tabular}
    \caption{The same as Table~\ref{tab:LMC-CEP} but for RR Lyrae stars in the LMC. \label{tab:LMC-RRL}}
\end{table*}

\begin{table*}\hspace*{-0.6cm}
    \centering
\begin{tabular}{ccccccccc}\hline
                ID &         $M/$M$_\odot$ &      $\log R/$R$_\odot$ &    $\log L/$L$_\odot$ &    $T_{\text{eff}}/K$  & $M_I$ &    $M_V$ & $V-I$ &  $W$ \\\hline\hline
0001 & 0.595 $\pm$ 0.057 & 0.708 $\pm$ 0.020 & 1.661 $\pm$ 0.043 & 6640 $\pm$ 110 & 0.142 $\pm$ 0.089 & 0.65 $\pm$ 0.10 & 0.490 $\pm$ 0.029 & -0.613 $\pm$ 0.081 \\
0002 & 0.581 $\pm$ 0.054 & 0.709 $\pm$ 0.021 & 1.604 $\pm$ 0.043 & 6420 $\pm$ 100 & 0.277 $\pm$ 0.089 & 0.79 $\pm$ 0.10 & 0.532 $\pm$ 0.029 & -0.560 $\pm$ 0.081 \\
0003 & 0.648 $\pm$ 0.050 & 0.755 $\pm$ 0.019 & 1.723 $\pm$ 0.044 & 6570 $\pm$ 110 & -0.006 $\pm$ 0.091 & 0.48 $\pm$ 0.11 & 0.484 $\pm$ 0.030 & -0.754 $\pm$ 0.078 \\
0008 & 0.697 $\pm$ 0.050 & 0.767 $\pm$ 0.019 & 1.675 $\pm$ 0.042 & 6270 $\pm$ 100 & 0.096 $\pm$ 0.087 & 0.657 $\pm$ 0.099 & 0.569 $\pm$ 0.028 & -0.791 $\pm$ 0.077 \\
0009 & 0.683 $\pm$ 0.053 & 0.785 $\pm$ 0.020 & 1.769 $\pm$ 0.042 & 6470 $\pm$ 110 & -0.110 $\pm$ 0.088 & 0.418 $\pm$ 0.099 & 0.529 $\pm$ 0.029 & -0.933 $\pm$ 0.083 \\
0011 & 0.669 $\pm$ 0.051 & 0.742 $\pm$ 0.020 & 1.697 $\pm$ 0.042 & 6530 $\pm$ 100 & 0.097 $\pm$ 0.087 & 0.601 $\pm$ 0.099 & 0.508 $\pm$ 0.029 & -0.695 $\pm$ 0.079 \\
0012 & 0.689 $\pm$ 0.051 & 0.758 $\pm$ 0.019 & 1.709 $\pm$ 0.042 & 6450 $\pm$ 100 & 0.049 $\pm$ 0.087 & 0.571 $\pm$ 0.098 & 0.527 $\pm$ 0.028 & -0.774 $\pm$ 0.078 \\
0015 & 0.637 $\pm$ 0.058 & 0.730 $\pm$ 0.021 & 1.687 $\pm$ 0.044 & 6570 $\pm$ 110 & 0.106 $\pm$ 0.090 & 0.59 $\pm$ 0.10 & 0.500 $\pm$ 0.030 & -0.672 $\pm$ 0.084 \\
0019 & 0.604 $\pm$ 0.054 & 0.751 $\pm$ 0.021 & 1.678 $\pm$ 0.044 & 6410 $\pm$ 100 & 0.092 $\pm$ 0.092 & 0.62 $\pm$ 0.10 & 0.540 $\pm$ 0.029 & -0.750 $\pm$ 0.083 \\
0023 & 0.644 $\pm$ 0.058 & 0.739 $\pm$ 0.022 & 1.685 $\pm$ 0.044 & 6510 $\pm$ 110 & 0.098 $\pm$ 0.091 & 0.63 $\pm$ 0.10 & 0.518 $\pm$ 0.029 & -0.700 $\pm$ 0.085 \\\hline
\end{tabular}
    \caption{The same as Table~\ref{tab:LMC-CEP} but for RR Lyrae stars in the SMC. \label{tab:SMC-RRL}}
\end{table*}

\begin{table*}\hspace*{-0.6cm}
    \centering
\begin{tabular}{ccccccccc}\hline
                ID &         $M/$M$_\odot$ &      $\log R/$R$_\odot$ &    $\log L/$L$_\odot$ &    $T_{\text{eff}}/K$  & $M_I$ &    $M_V$ & $V-I$ &  $W$ \\\hline\hline
00197 & 0.728 $\pm$ 0.055 & 0.756 $\pm$ 0.020 & 1.661 $\pm$ 0.043 & 6300 $\pm$ 100 & 0.136 $\pm$ 0.089 & 0.70 $\pm$ 0.10 & 0.570 $\pm$ 0.029 & -0.742 $\pm$ 0.082 \\
00207 & 0.567 $\pm$ 0.052 & 0.669 $\pm$ 0.019 & 1.603 $\pm$ 0.042 & 6730 $\pm$ 100 & 0.308 $\pm$ 0.088 & 0.78 $\pm$ 0.10 & 0.470 $\pm$ 0.029 & -0.423 $\pm$ 0.078 \\
00236 & 0.579 $\pm$ 0.050 & 0.717 $\pm$ 0.019 & 1.722 $\pm$ 0.042 & 6810 $\pm$ 100 & 0.044 $\pm$ 0.087 & 0.500 $\pm$ 0.099 & 0.450 $\pm$ 0.029 & -0.656 $\pm$ 0.078 \\
00242 & 0.630 $\pm$ 0.055 & 0.718 $\pm$ 0.020 & 1.621 $\pm$ 0.043 & 6410 $\pm$ 100 & 0.233 $\pm$ 0.089 & 0.77 $\pm$ 0.10 & 0.539 $\pm$ 0.029 & -0.609 $\pm$ 0.080 \\
00254 & 0.539 $\pm$ 0.053 & 0.675 $\pm$ 0.020 & 1.605 $\pm$ 0.043 & 6700 $\pm$ 110 & 0.283 $\pm$ 0.089 & 0.77 $\pm$ 0.10 & 0.478 $\pm$ 0.029 & -0.456 $\pm$ 0.080 \\
00276 & 0.645 $\pm$ 0.058 & 0.705 $\pm$ 0.021 & 1.654 $\pm$ 0.043 & 6630 $\pm$ 100 & 0.195 $\pm$ 0.089 & 0.67 $\pm$ 0.10 & 0.488 $\pm$ 0.029 & -0.564 $\pm$ 0.083 \\
00397 & 0.569 $\pm$ 0.054 & 0.679 $\pm$ 0.020 & 1.633 $\pm$ 0.042 & 6760 $\pm$ 100 & 0.243 $\pm$ 0.088 & 0.710 $\pm$ 0.099 & 0.461 $\pm$ 0.029 & -0.473 $\pm$ 0.080 \\
00409 & 0.691 $\pm$ 0.053 & 0.757 $\pm$ 0.020 & 1.691 $\pm$ 0.043 & 6380 $\pm$ 100 & 0.080 $\pm$ 0.089 & 0.60 $\pm$ 0.10 & 0.540 $\pm$ 0.029 & -0.763 $\pm$ 0.080 \\
00414 & 0.643 $\pm$ 0.059 & 0.713 $\pm$ 0.022 & 1.640 $\pm$ 0.044 & 6520 $\pm$ 110 & 0.214 $\pm$ 0.091 & 0.72 $\pm$ 0.10 & 0.515 $\pm$ 0.029 & -0.590 $\pm$ 0.084 \\
00418 & 0.656 $\pm$ 0.051 & 0.728 $\pm$ 0.020 & 1.617 $\pm$ 0.042 & 6350 $\pm$ 100 & 0.260 $\pm$ 0.089 & 0.79 $\pm$ 0.10 & 0.550 $\pm$ 0.028 & -0.592 $\pm$ 0.079 \\\hline
\end{tabular}
    \caption{The same as Table~\ref{tab:LMC-CEP} but for RR Lyrae stars in the Galactic Bulge. \label{tab:BLG-RRL}}
\end{table*}

\section{Conclusions}
Our interest is in determining global stellar parameters of Cepheids and RR~Lyraes, such as $M,L,T_{\text{eff}}$ and $R$ and in constraining input stellar physics. 
The period--mean density relation suggests that the observable period is the most important quantity in constraining the global stellar parameters. 
However, in this study, using recently published pulsation models for Cepheids and RR~Lyraes, we have used modern data analysis methods to demonstrate that light curve structure plays a statistically significant part in determining these parameters. 
Using these techniques we have developed a catalogue of predicted masses, luminosities, radii and effective temperatures for OGLE Cepheids and RR~Lyraes. 
We have furthermore used the predicted luminosities and hence predicted $V$- and $I$-band theoretical magnitudes to construct a Wesenheit function and estimate the distance to the LMC and SMC. 
Our calculated distances are statistically consistent with the latest estimates from eclipsing binaries.

Certainly our results are model-dependent. 
If there are systematic errors in the models or missing physics this will be reflected in the final results. 
However this should be then viewed as an incentive to improve or further constrain stellar pulsation models. 
In the future, it will be interesting to compare the estimates made using this method to independent measures from other methods, such as interferometric radii, Gaia luminosities, and dynamical masses. 

Our results could be improved by a more extensive or more uniform input model grid. 
For example, we are currently able to apply this method to only about 10\% of the Cepheids observed in the LMC due to the relatively limited parameter space spanned by the grid of models. 
Furthermore, some unpropagated systematic errors exist in the resulting estimated parameters due to some theoretical parameters being held fixed, such as using only one composition for all Cepheid models. 
We plan follow-up studies in which we compute a more extensive grid in order to obtain more accurate estimates of these parameters.

\section*{Acknowledgements}
We thank J{\o}rgen Christensen-Dalsgaard and the anonymous referee for comments and suggestions which improved the manuscript. 
Funding for the Stellar Astrophysics Centre is provided by The Danish National Research Foundation (Grant agreement no.: DNRF106). SMK thanks the Stellar Astrophysics Centre (SAC), Aarhus University, SUNY Oswego and the Indo-US Science and Technology Forum for supporting a visit to SAC in July 2019 where much of this work was completed. 
AB acknowledges research grant \#11850410434 awarded by the National Natural Science Foundation of China through the Research Fund for International Young Scientists, and a China Postdoctoral General Grant (2018M640018).

\bibliographystyle{mnras}
\bibliography{Bellinger}




\bsp	
\label{lastpage}
\end{document}